\documentstyle[preprint,eqsecnum,aps]{revtex}
\begin{document}
\draft
\everymath={\displaystyle}
\title{\bf What $\pi-\pi$ scattering tells us about\\ chiral
perturbation theory}
\author{J. Stern and H. Sazdjian}
\address{
Division de Physique Th\'eorique\thanks{Unit\'{e} de Recherche des
Universit\'{e}s Paris 11 et Paris 6 associ\'{e}e au CNRS},
Institut de Physique Nucl\'eaire, \\ F-91406 Orsay Cedex, France
}
\author{N. H. Fuchs}
\address{
Department of Physics, Purdue University, West Lafayette IN 47907 \\
\vspace{12cm}
}

\preprint{PURD-TH-93-01}
\maketitle
{\flushleft IPNO/TH 92-106}
\pagebreak
\begin{abstract}
We describe a rearrangement of the standard expansion of the symmetry
breaking part of the QCD effective Lagrangian that includes into each
order additional terms which in the standard chiral perturbation
theory ($\chi$PT) are relegated to higher orders.  The new expansion
represents a systematic and unambiguous generalization of the standard
$\chi$PT, and is more likely to converge rapidly.  It provides a
consistent framework for a measurement of the importance of additional
``higher order'' terms whose smallness is usually assumed but has never
been checked.  A method of measuring, among other quantities, the QCD
parameters $\hat{m}\langle\bar{q}q\rangle$ and the quark mass ratio
$m_s/\hat{m}$ is elaborated in detail.  The method is illustrated
using various sets of available data.  Both of these parameters might
be considerably smaller than their respective leading order standard
$\chi$PT values.  The importance of new, more accurate, experimental
information on low-energy $\pi-\pi$ scattering is stressed.
\end{abstract}

\narrowtext
\section {\bf Introduction}

Owing to quark confinement, the connection between QCD correlation
functions and hadronic observables is far from being straightforward.
In the low-energy domain, such a connection is described by chiral
perturbation theory ($\chi$PT) \cite{sw79,gl84,gl85}. The latter provides
a complete parametrization (in terms of an effective Lagrangian) of
low-energy off-shell correlation functions of quark bilinears,which
should take into account: (i) the normal and anomalous Ward identities
of chiral symmetry, explicitly broken by quark masses; (ii)
spontaneous breakdown of chiral symmetry; (iii) analyticity, unitarity
and crossing symmetry.  On the other hand, such a parametrization
(effective Lagrangian) should be sufficiently general, and should not
introduce any additional dynamical assumptions beyond those listed
above, that could be hard to identify as emerging from QCD.  The
specificity of QCD then resides in numerical values of low-energy
constants which characterize the above parametrization. The
theoretical challenge is to calculate these low-energy parameters from
the fundamental QCD Lagrangian.  While such a calculation is awaited,
these parameters can be subjected to experimental investigation.
Chiral symmetry guarantees that the same parameters that are
introduced through the low-energy expansion of QCD correlation
functions also define the low-energy expansion of hadronic observables
--- pseudoscalar meson masses, transition and scattering amplitudes.

In this paper, a new method will be elaborated that allows a detailed
measurement of certain low-energy parameters, using the $\pi-\pi$
elastic scattering data \cite{gl84,drv}. Instead of concentrating on a
particular set of scattering lengths and effective ranges \cite{gl84}
whose extraction from experimental data is neither easy nor accurate,
emphasis will be put on a detailed fit of the scattering amplitude in
a whole low-energy domain of the Mandelstam plane, including the
unphysical region.  In this way it is possible to obtain some {\it
experimental} insight on the low-energy parameter $2\hat{m}B_0$, where
$\hat{m}$ is the average of the up and down quark masses, $B_0$ is the
condensate
\begin{equation}
B_0 = - \frac{1}{F_0^2} \langle 0|\bar{u}u|0\rangle = -
\frac{1}{F_0^2} \langle 0|\bar{d}d|0\rangle  = - \frac{1}{F_0^2}
\langle  0|\bar{s}s|0\rangle
\end{equation}
and $|0\rangle $ and $F_0$ stand for the ground state and pion decay
constant respectively at $m_u = m_d = m_s = 0$.  It is usually {\it
assumed} that the parameter $2\hat{m} B_0$ differs from the pion mass
squared by not more than $(1-2)\%$ \cite{gl85}, and the standard
chiral perturbation theory could hardly tolerate an important
violation of this assumption \cite{gor-gw}. On the other hand, this
assumption has never been confronted with experiment otherwise than
indirectly -- through the Gell-Mann Okubo formula for pseudoscalar
meson masses \cite{gor-gw}. However, even the latter represents at
best a consistency argument rather than a proof: the Gell-Mann Okubo
formula can hold quite independently of the relation between
$2\hat{m}B_0$ and $M_\pi^2$ \cite{fss91}.  An independent measurement
of $2\hat{m}B_0$ is not only possible (as shown in the present work)
but, for several reasons, it appears to be desirable:

(i) The effective Lagrangian ${\cal L}_{eff}$ contains, in
principle, an infinite number of low-energy constants, which are all
related to (gauge invariant) correlation functions of massless QCD.
Among them, $B_0$ plays a favored role:  The order of magnitude of all
low-energy constants other than $B_0$ can be estimated using sum-rule
techniques \cite{sumrules}, which naturally bring in the scale
$\Lambda \sim 1~GeV$ characteristic of massive bound states.  The
expected order of magnitude of a low-energy constant related to a
connected $N$-point ($N > 1$) function of quark bilinears
$\bar{q}\Gamma q$, that is not suppressed by the Zweig rule or by a
symmetry, is $F_0^2 \Lambda^{2-N}$ multiplied by a dimensionless
constant of order 1.  If quarks were not confined \cite{njl}, a
similar estimate would relate $B_0$ and the mass of asymptotic fermion
states with quark quantum numbers.  However, in a confining theory, no
similar relation between $B_0$ and the spectrum of massive bound
states can be derived:  $\bar{q}q$ is an irreducible color singlet and
there is no complete set of intermediate states which could be
inserted into the matrix element $\langle 0| \bar{q}q |0\rangle$.
$B_0$ could be as large as $\Lambda \sim 1~GeV$ or as small as the
fundamental order parameter of chiral symmetry breaking, $F_0 \sim
90~MeV$.  {\it A priori}, there is no way to decide in favor of one of
these scales, at least before the non-perturbative sector of QCD is
controlled analytically or by reliable numerical methods, using, for
instance, sufficiently large lattices.  In this paper, we suggest how
the question of the scale of $B_0$ can be addressed experimentally.

(ii) To the extent that $2\hat{m}B_0$ and $(m_s + \hat{m})B_0$
are close to $M_\pi^2$ and $M_K^2$, respectively, the ratio of quark
masses
\begin{equation}
r \equiv m_s/\hat{m}
\end{equation}
must approach $2 \frac{M_K^2}{M_\pi^2} -1 = 25.9$ \cite{gor-gw,w-gl}.
There exists an independent measurement of the ratio $r$ in terms of
observed deviations from the Goldberger-Treiman relation \cite{fss90}
in non-strange and strange baryon channels.  This model-independent
measurement indicates a considerably lower value for $r$ than 25.9,
unless the pion-nucleon coupling constant turns out to be below the
value given by Koch and Pietarinen \cite{koch-piet} by at least 4--5
standard deviations \cite{arndt-etal}.

(iii)  A reformulation of $\chi$PT which allows $2\hat{m}B_0$ to be
considerably lower than $M_\pi^2$ has been given in Ref.\
\cite{fss91}.  It is as systematic and unambiguous as the standard
$\chi$PT itself, and is particularly suitable in the case where $B_0$ is
as small as $F_\pi$.  It is based on a different expansion of the same
effective Lagrangian, with the same infinity of independent terms.  To
all orders, the two perturbative schemes are identical but, in each
finite order, they can (but need not) substantially differ.  For each
given order, the new scheme contains more parameters than the standard
$\chi$PT, the latter being reproduced for special values of these
additional parameters.  Already at the leading order $O(p^2)$, the new
scheme contains one additional free parameter
\begin{equation}
\eta = \frac{2\hat{m}B_0}{M_\pi^2}.
\end{equation}
If $\eta$ is set equal to 1, one recovers the leading $O(p^2)$ order
of the standard $\chi$PT.  The new expansion can therefore be formally
viewed as a {\it generalization} of the standard scheme and -- in this
sense -- it will be referred to as {\it improved $\chi$PT}, since it aims
to improve the convergence of the standard perturbation theory.
Demonstrating that such an improvement is irrelevant, by measuring,
for instance, the ratio (1.3) and finding it close to unity, would be
an important experimental argument in favor of the standard $\chi$PT.

(iv) In some cases, the convergence of standard $\chi$PT actually appears
to be rather slow.  Most of the indications in this direction can be
traced back to the fact that the leading $O(p^2)$ order of the
standard $\chi$PT underestimates the Goldstone boson interaction and, in
particular, the $\pi-\pi$ scattering amplitude.  This manifests itself
through virtual processes and/or final state interactions, as in
$\gamma \gamma \rightarrow \pi^0 \pi^0$ \cite{gammagamma}, $\eta-3\pi$
\cite{eta3pi}, etc.  It might even be that although the next order
$O(p^4)$ improves the situation, it fails to reach the precision we
may rightly expect from it.  For example, the $I=0$ $s$ wave $\pi-\pi$
scattering length, which is $a_0^0 = 0.16$ in leading order
\cite{sw66}, gets shifted to $a_0^0 = 0.20$ by $O(p^4)$ corrections
\cite{gl84}, while the ``experimental value''
\cite{yellow,fp,nagels} is $a_0^0 = 0.26 \pm 0.05$.  (In this paper it
will be argued that scattering lengths are not the best quantities to
look at.  A more detailed amplitude analysis will reveal a possible
amplification of the discrepancy, which exceeds one standard
deviation.)

(v) $\chi$PT should be merely viewed as a theoretical framework for a
precise measurement of low-energy QCD correlation functions.  Its
predictive power rapidly decreases with increasing order in the chiral
expansion : More new parameters enter at each order and more
experimental data have to be included to pin them down.  For this
reason, a slow convergence rate might sometimes lead to a
qualitatively wrong conclusion with respect to a measurement based
only on the first few orders. This might concern, in particular, the
measurement of the ratio $\eta$ (1.3) within the standard $\chi$PT. In the
corresponding leading order, $\eta$ is fixed to be 1, independently of
any experimental data.  This property of standard $\chi$PT could bias the
measurement of $\eta$ if $\eta$ turned out to be considerably
different from 1: one would presumably have to go to a rather high
order and include a large set of data to discover the truth.  In this
case, the {\it improved $\chi$PT} would be a more suitable framework to
measure $\eta$ faithfully.  The reason is that in the improved $\chi$PT,
$\eta$ is a {\it free parameter} from the start: It defines the {\it
leading order} $\pi-\pi$ amplitude.  Neglecting, for simplicity,
Zweig-rule violation (cf. Ref.\ \cite{fss91} and Sec. IV A), the
latter reads
\begin{equation}
A(s|tu) = \frac{1}{F_0^2} (s-\eta M_\pi^2).
\end{equation}
Using in this formula the value of $a_0^0 = 0.26 \pm 0.05$,
one concludes that $\eta = 0.4 \pm 0.4$ already at the leading order.
The measurement then has more chances to saturate rapidly -- say, at
the one-loop level -- provided $\eta$ is much closer to $0.4 \pm 0.4$
than to 1.  The same remark applies to measurements of the quark mass
ratio $r = m_s/\hat{m}$, which, incidentally, is closely related to
$\eta$ \cite{fss90}: A slow convergence of the standard $\chi$PT could
lower the leading-order result $r = 25.9$ by considerably more than
the usually quoted $(10-20)\%$ \cite{gl85,donoghuereview}.

(vi) The question of the actual value of $\eta$ and/or of $r =
m_s/\hat{m}$ has to be settled experimentally.  None of the known
properties of QCD, nor the fact that light quark masses are tiny
compared with the hadronic scale $\Lambda \sim 1~GeV$, imply that
$\eta$ should be close to 1 and that $r$ should be close to 25.9.  The
proof of this negative statement is provided by the existence of a
mathematically consistent generalization of the standard $\chi$PT that
does not contradict any known fundamental property of QCD and allows
for any value of $\eta$ between 0 and 1 (and for any value of $r$
between 6.3 and 25.9) \cite{fss91}.  Only in the special case of
$\eta$ and $r$ close to 1 and 25.9, respectively, can the standard
$\chi$PT claim a decent rate of convergence.

In Sec. II, the precise mathematical definition of the improved
$\chi$PT, in terms of the effective Lagrangian, is briefly summarized.
It is not a purpose of this paper to present a full formal development
of this theory; incidentally, most of it can be read off from existing
calculations \cite{gl85} after rather minor extensions (which will be
presented elsewhere).  Here, we will mainly concentrate on
phenomenological aspects of the problem in connection with low-energy
$\pi-\pi$ scattering.  The content of Sec. III is independent of any
particular $\chi$PT scheme.  In that section, a new low-energy
representation of the $\pi-\pi$ scattering amplitude is given that
provides the most general solution of analyticity, crossing symmetry
and unitarity up to and including the chiral order $O(p^6)$.  (Partial
wave projections of this representation coincide with a particular
truncation of the well-known Roy equations \cite{roy}.)  Subsequently,
this representation is used both to constrain the experimental data
and to perform a comparison with theoretical amplitudes as predicted
by the two versions of $\chi$PT.  For the case of the improved
$\chi$PT, the one-loop amplitude is worked out in Sec. IV.  Finally, a
method permitting a detailed fit of the experimental amplitude in a
whole low-energy domain of the Mandelstam plane is developed in Sec.
V.  This method is then applied to various sets of existing data.

\section{\bf Formulation of improved $\chi$PT}

Following rather closely the off-shell formalism which was elaborated
some time ago by Gasser and Leutwyler \cite{gl85}, we consider the
generating functional $Z(v^\mu,a^\mu,\chi)$ of connected Green
functions made up from $SU(3)\times SU(3)$ vector and axial currents as
well as from scalar and pseudoscalar quark densities, as defined in
QCD with three massless flavors.  The sources $v^\mu,~a^\mu$ and
$\chi$ are specified through the Lagrangian
\begin{equation}
{\cal L} = {\cal L}_{QCD} + \bar{q}(\not \! v + \not \! a \gamma_5)q -
\bar{q}_R \chi q_L - \bar{q}_L \chi^{\dag} q_R,
\end{equation}
which defines the vacuum-to-vacuum amplitude $\exp i Z$.  Here,
$q_{L,R} = \case{1}{2}(1 \mp \gamma_5)q$ stand for the light quark
fields $u,\,d,\,s$ and ${\cal L}_{QCD}$ is invariant under global
$SU(3) \times SU(3)$ transformations of $q_L$ and $q_R$.  $v^\mu$ and
$a^\mu$ are traceless and hermitean, whereas
\begin{equation}
\chi = s + ip
\end{equation}
is a general $3 \times 3$ complex matrix ($s$ and $p$ are hermitean).
Explicit chiral symmetry breaking by quark masses is accounted for by
expanding $Z$ around the point
\begin{equation}
v^\mu = a^\mu = 0,~~~~~\chi = {\cal M}_q \equiv \left(
\begin{array}{rrr}
m_u &~~  &~~  \\
{}~~ & m_d &~~ \\
{}~~ &~~  & m_s
\end{array}
\right).
\end{equation}
The scalar-pseudoscalar source $\chi$ and the quark mass matrix ${\cal
M}_q$ are closely tied together by chiral symmetry.  (Notice that our
source $\chi$ differs from the $\chi$ defined in Ref.\ \cite{gl85} by
a factor of $2B_0$.)

Instead of calculating $Z$, the effective theory parametrizes it by
means of an effective Lagrangian which depends on the sources and on
eight Goldstone boson fields
\begin{equation}
U(x) = \exp \frac{i}{F_0} \sum_{a=1}^8 \lambda^a \varphi_a(x).
\end{equation}
Leaving aside anomaly contributions described by the Wess-Zumino
action, the effective Lagrangian ${\cal L}_{eff}(U,v^\mu,a^\mu,\chi)$
is merely restricted by the usual space-time symmetries and by the
requirement of invariance under local chiral transformations
[$\Omega_{L,R} \in SU(3)$]
\begin{equation}
U(x) \rightarrow \Omega_R(x) U(x) \Omega_L^{\dag}(x),~~~\chi(x)
\rightarrow \Omega_R(x) \chi(x) \Omega_L^{\dag}(x)
\end{equation}
compensated by the inhomogeneous transformation of the sources $v^\mu$
and $a^\mu$:
\begin{eqnarray}
v^\mu + a^\mu &\rightarrow& \Omega_R (v^\mu + a^\mu + i \partial ^\mu)
\Omega_R^{\dag} \nonumber \\
v^\mu - a^\mu &\rightarrow& \Omega_L (v^\mu - a^\mu + i \partial ^\mu)
\Omega_L^{\dag}.
\end{eqnarray}
(This gauge invariance of the nonanomalous part of $Z$ is necessary
and sufficient to reproduce all $SU(3) \times SU(3)$ Ward identities.)
Otherwise, the effective Lagrangian remains unrestricted.

${\cal L}_{eff}$ can be written as an infinite series of local terms,
\begin{equation}
{\cal L}_{eff} = \sum_{n,m} \ell ^{nm} {\cal L}_{nm},
\end{equation}
where ${\cal L}_{nm}$ denotes an invariant under the transformations
(2.5) and (2.6) that contains the $n$-th power of the covariant
derivatives $D_\mu$ and the $m$-th power of the scalar-pseudoscalar
source $\chi$.  The sum over independent invariants that belong to the
same pair of indices $(n,m)$ is understood.  The covariant derivatives
are defined as
\begin{equation}
D_\mu U = \partial _\mu U - i(v_\mu + a_\mu)U + iU(v_\mu - a_\mu),
\end{equation}
and likewise for $D_\mu \chi$.  The expansion coefficients $\ell
^{nm}$ represent properly subtracted linear combinations of massless
QCD correlation functions that involve $n$ vector and/or axial
currents and $m$ scalar and/or pseudoscalar densities, all taken at
vanishing external momenta.  The first two terms in the sum (2.7), for
instance, read ($n$ is even)
\begin{eqnarray}
\ell ^{01} {\cal L}_{01} &=& \frac{1}{2} F_0^2 B_0 \langle U^{\dag}\chi
+ \chi^{\dag}U \rangle \nonumber \\
\ell ^{20} {\cal L}_{20} &=& \frac{1}{4} F_0^2 \langle D^\mu U^{\dag}
D_\mu U \rangle.
\end{eqnarray}
Everything said so far is rather general and independent of any
particular perturbative scheme.

Chiral perturbation theory is an attempt to reorder the infinite sum
(2.7) as
\begin{equation}
{\cal L}_{eff} = \sum_d {\cal L}^{(d)},
\end{equation}
where ${\cal L}^{(d)}$ collects all terms that in the limit
\begin{equation}
p \rightarrow 0,~~~M_\pi \rightarrow 0,~~~ p^2/M_\pi^2 ~~\text{fixed}
\end{equation}
behave as $O(p^d)$ ($p$ stands for external momenta).  In order to
relate the expansions (2.10) and (2.7), one needs to know the {\it
effective infrared dimension $d(m_q)$ of the quark mass}.  The
invariant ${\cal L}_{nm}$ then contributes as $O(p^{d_{nm}})$,  where
\begin{equation}
d_{nm} = n + m\,d(m_q).
\end{equation}
For infinitesimally small quark masses, one should have
\begin{equation}
d(m_q) = 2,~~~m_q \rightarrow 0.
\end{equation}
This follows from the mathematical fact that in QCD
\begin{equation}
\lim_{m_q \rightarrow 0} \frac{(m_i + m_j)B_0}{M_P^2} = 1.
\end{equation}
(Here, $i,j = u,d,s$ and $M_P$ is the mass of the pseudoscalar meson
$\bar{\imath}j,~i\not =j$.)  The assumption that in the {\it real
world, i.e.}, for physical values of quark masses, the effective
dimension of the quark mass is 2, underlies the {\it standard
$\chi$PT}.  It amounts to the well-known rule which asserts that each
insertion of the quark mass matrix and/or of the scalar-pseudoscalar
source $\chi$ counts as two powers of external momenta.  Equivalently,
the standard $\chi$PT can be viewed as an expansion around the limit
\begin{equation}
(p,m_q) \rightarrow 0,~~~p^2/M_P^2~~\text{fixed}.
\end{equation}
Since, by definition, the low-energy constants $\ell^{nm}$ are
independent of quark masses, they are $O(1)$ in the limit (2.15).

It is easy to see that the convergence of the standard $\chi$PT could
be seriously disturbed if $B_0 \ll \Lambda \sim 1~GeV$, say $B_0 \sim
100 ~MeV$ \cite{fss91,fss90}.  The expansion of $M_P^2$ reads $(i,j =
u,d,s;~ i \not = j)$
\begin{equation}
M_P^2 = (m_i + m_j)B_0 + (m_i + m_j)^2 A_0 + \ldots,
\end{equation}
where the dots stand for non-analytic terms and for higher order
terms.  $A_0$ can be expressed in terms of two-point functions of
scalar and pseudoscalar quark densities divided by $F_0^2$
\cite{fss90}.  It satisfies a superconvergent dispersion relation,
whose saturation leads to the order of magnitude estimate $A_0 \sim 1
- 5$.  For $B_0$ as small as 100 MeV, the first and second order terms
in Eq. (2.16) then become comparable for quark masses as small as (10
-- 50) MeV.  In order to accommodate this possibility, the {\it
improved $\chi$PT} attributes to the quark mass {\it and} to the
vacuum condensate parameter $B_0$ the effective dimension 1,
\begin{equation}
d(m_q) = d(B_0) = 1,
\end{equation}
reflecting their smallness compared to the scale $\Lambda$.  This does
not contradict mathematical statements such as (2.14).  It only means
that, although for physical values of quark masses the ratio in Eq.
(2.14) remains on the order of 1, it is allowed to differ from 1
considerably.

To summarize, in the improved $\chi$PT each insertion of the quark
mass-matrix ${\cal M}_q$ and/or of the scalar-pseudoscalar source
$\chi$ counts as a single power of external momentum (pion mass) {\it
and so does the parameter $B_0$}.  This leads to a new expansion of
the effective Lagrangian
\begin{equation}
{\cal L}_{eff} = \sum_d \tilde{\cal L}^d,
\end{equation}
where each $\tilde{\cal L}^d$ contains more terms ${\cal L}_{nm}$ than
does the corresponding term ${\cal L}^d$ in the case of the standard
counting.  The improved $\chi$PT is a simultaneous expansion in
$p/\Lambda, m_q/\Lambda$ and $B_0/\Lambda$ around the limit
\begin{equation}
(p,m_q,B_0) \rightarrow 0,~~~~ p^2/M_P^2 ~\text{and}~
m_qB_0/M_P^2~~\text{fixed}.
\end{equation}
This is just another way to realize the chiral limit (2.11).
The fact that -- in the effective theory -- we treat $B_0$ as an
arbitrary expansion parameter does not contradict the general belief
that, within QCD, this parameter is fixed and -- hopefully --
calculable.  After all, quantum electrodynamics is also based on an
expansion in $\alpha$, in spite of the general belief that there might
exist a more fundamental theory in which the value of $\alpha$ is
fixed and calculable \cite{edd}.
\section{\bf Reconstruction of the low-energy \protect \\
$\pi-\pi$ scattering amplitude \protect \\
neglecting $O(\lowercase{p}^8)$ effects.}

The analysis of low-energy $\pi-\pi$ scattering, traditionally based
on analyticity, crossing symmetry and unitarity
\cite{chew-mandelstam,yellow,roy}, considerably simplifies if, in
addition, one takes into account the Goldstone character of the pion.
First, in the chiral limit (2.11), higher $(\ell \geq 2)$ partial
waves are suppressed.  The reason stems from the fact that in the
limit (2.11) the whole amplitude behaves as $O(p^2)$ and furthermore,
it does not contain light dipion bound state poles.  Unitarity then
implies that the scattering amplitude is dominantly real, since its
imaginary part behaves as $O(p^4)$.  Analyticity then forces the
leading $O(p^2)$ part of the amplitude $A(s|tu)$ to be a polynomial in
the Mandelstam variables.  Furthermore, higher than first order
polynomials are excluded: They would be $O(p^2)$ only provided their
coefficients blew up as $M_\pi^2 \rightarrow 0$, which would
contradict the finiteness of the S-matrix in the limit $m_q
\rightarrow 0$ with the external momenta kept fixed at a
non-exceptional value.  Finally, crossing symmetry allows one to
express the $O(p^2)$ part of the scattering amplitude $A(s|tu)$ as
\begin{equation}
A_{Lead}(s|tu) = \frac{\alpha}{3F_\pi^2}M_\pi^2 +
\frac{\beta}{3F_\pi^2} (3s-4M_\pi^2),
\end{equation}
where $\alpha$, $\beta$ are two dimensionless constants which are
$O(1)$ in the chiral limit.  The linear amplitude (3.1) does not
contribute to $\ell \geq 2$ partial waves.  Consequently, the latter
behave in the chiral limit as $O(p^4)$ and, owing to unitarity, the
absorptive parts of $\ell \geq 2$ waves are suppressed at least to
$O(p^8)$. This conclusion holds independently of more quantitative
predictions of $\chi$PT, which in the actual case merely concern
the values of the two parameters ${\alpha}$ and ${\beta}$
in Eq.~(3.1).

The second simplification resides in the suppression of inelasticities
arising from intermediate states that consist of more than two Goldstone
bosons.  The behavior of $n$-pion invariant phase space in the chiral
limit (2.11) is given by its dimension: It scales like $p^{2n-4}$.
Amplitudes with an arbitrary number of external pion legs are
dominantly $O(p^2)$.  Consequently, the contribution of multi-pion $(n
> 2)$ intermediate states to the absorptive part of the elastic $\pi-\pi$
amplitude is suppressed in the chiral limit at least to $O(p^8)$.

The smallness of higher partial waves and of inelasticities are of
course well-known phenomenological facts \cite{yellow}.  It
is important that these ``remarkable accidents" (see page 53 of
\cite{yellow}) can be put under the rigorous control of chiral power
counting: The previous discussion suggests that a rather simple
amplitude analysis of low-energy $\pi-\pi$ scattering can be performed
{\it up to and including $O(p^6)$ contributions}.  In the following we
confirm and elaborate this expectation in detail.  It will be shown in
particular that, neglecting $O(p^8)$ contributions, the whole
scattering amplitude can be expressed in terms of low-energy $s$ and
$p$ wave phase shifts and six (subtraction) constants.  (The latter
are related to the experimental phase shifts $via$ unitarity.) The
resulting expression (3.2) will prove particularly useful both for
constraining low-energy experimental data and for providing a basis
for a confrontation of chiral perturbation theory up to two loops with
experiment.

\subsection{\bf Statement of the theorem}

Let $\Lambda$ denote a scale (slightly) below the threshold for
production of non-Goldstone particles.  The $\pi-\pi$ amplitude can be
written as
\begin{eqnarray}
\frac{3}{32\pi} A(s|tu) &=& T(s)+ T(t) + T(u) \nonumber \\
&+& \frac{1}{3} [2U(s) - U(t) - U(u)]
\nonumber \\
&+& \frac{1}{3}[(s-t) V(u) + (s-u)V(t)] \nonumber \\
&+& R_{\Lambda}(s|t,u).
\end{eqnarray}
The remainder, $R_{\Lambda}$, behaves in the chiral limit as $O(p^8)$
relative to the scale $\Lambda$: up to possible logarithmic terms,
\begin{equation}
R_{\Lambda} = O([p/\Lambda]^8),
\end{equation}
where $p$ stands for external pion momenta.  In practice, $\Lambda
\lesssim 1~GeV$.  The functions $T,U$ and $V$ are analytic for $s <
4M_\pi^2$, whereas for $4M_\pi^2 < s < \Lambda^2$ their
discontinuities are given by the three lowest partial wave amplitudes
$f_\ell^I(s)$:\footnote{Notation and normalization are reviewed in
Appendix A.}
\begin{eqnarray}
I\!m\,T(s) &=& \frac{1}{3} \{ I\!m\,f_0^0(s) + 2I\!m\,f_0^2(s) \}
\nonumber \\
I\!m\,U(s) &=& \frac{1}{2} \{ 2I\!m\,f_0^0(s) - 5I\!m\,f_0^2(s) \}
\nonumber \\
I\!m\,V(s) &=& \frac{27}{2} \frac{1}{s-4M_\pi^2}I\!m\,f_1^1(s).
\end{eqnarray}
The real parts of the functions $T,U$ and $V$ are defined only up to
polynomials
\begin{eqnarray}
\delta T(s) &=& x(s- \frac{4}{3} M_\pi^2) \nonumber \\
\delta U(s) &=& y_0 + y_1 s+ y_2 s^2 + y_3s^3 \nonumber \\
\delta V(s) &=& - (y_1 + 4M_\pi^2 y_2 + 16M_\pi^4 y_3) + (y_2 +
12M_\pi^2 y_3) s -3 y_3 s^2 ,
\end{eqnarray}
where $x$ and the $y$'s are five arbitrary real constants: because of
the relation $s+t+u=4M_\pi^2$, the two sets of amplitudes $T,U,V$ and
$T+\delta T, U+\delta U, V+\delta V$ lead to the same scattering
amplitude A.  (It is shown in Appendix B that Eqs.~(3.5) actually
represent the most general transformation of $T,U,V$ leaving the
scattering amplitude invariant.)  After conveniently fixing the
``gauge freedom'' (3.5), the functions $T,U$ and $V$ can be written as

\begin{eqnarray}
T(s) =& t_0 + t_2 s^2 + t_3 s^3 +
& \frac{s^3}{\pi} \int_{4M_\pi^2}^{{\Lambda}^2}
\frac{dx}{x^3} \frac{1}{x-s}I\!m\,T(x) \nonumber \\
U(s) =&&\frac{s^3}{\pi} \int_{4M_\pi^2}^{{\Lambda}^2}
\frac{dx}{x^3} \frac{1}{x-s} I\!m\,U(x) \nonumber \\
V(s) =& v_1 + v_2s + v_3s^2 + &\frac{s^2}{\pi}\int_{4M_\pi^2}^{{\Lambda}^2}
\frac{dx}{x^2}
\frac{1}{x-s} I\!m\,V(x),
\end{eqnarray}
where the imaginary parts are given by Eqs.~(3.4) and the $t$'s and
$v$'s are constants.  It will be shown shortly that Eq.~(3.2) is a
rigorous consequence of analyticity and crossing symmetry and of the
Goldstone nature of the pion.

\subsection{\bf Unitarity}

The low-energy representation (3.2) of the scattering amplitude is
exact up to an $O(p^8)$ remainder.  In the whole interval $4M_\pi^2
< s < \Lambda^2$, unitarity can be imposed with the same accuracy
in terms of partial waves $f_\ell^I$.  As already pointed out,
deviations from the unitarity condition \[
I\!m\,f_{\ell}^I(s) =
\sqrt{\frac{s-4M_\pi^2}{s}} |f_\ell^I(s)|^2
\]above the inelastic threshold are of the order $O(p^8)$.  The amplitude
(3.2) contains all partial waves.  For $\ell \geq 2$, the partial
waves are real.  Nevertheless, unitarity automatically is satisfied for
$\ell \geq 2$ up to $O(p^8)$ terms, since higher partial waves anyway
are $O(p^4)$ or smaller.  Consequently, it is sufficient to impose
unitarity for the three lowest waves $f_0^0, f_1^1$ and $f_0^2$
(hereafter denoted as $f_a, a = 0,1,2$ according to their isospin).
Projections of Eq.~(3.2) into the three lowest partial waves read
\begin{mathletters}
\begin{eqnarray}
Re\,f_a(s) &=& P_a(s) + \frac{s^3}{\pi}
{\int\hspace{-1em}-}_{4M_\pi^2}^{{\Lambda}^2}
\frac{dx}{x^3} \frac{I\!m\,f_a(x)}{x-s} \nonumber \\
&+& \frac{1}{\pi} \int_{4M_\pi^2}^{{\Lambda}^2} \frac{dx}{x}
\sum_{b=0}^2 W_{ab}(s,x) I\!m\,f_b(x) + O(p^8)
\end{eqnarray}
for the two $s$ waves $(a = 0,2)$, whereas the $p$ wave projection is
\begin{eqnarray}
Re\,f_1(s) &=& P_1(s) + \frac{s^2(s-4M_\pi^2)}{\pi}
{\int\hspace{-1em}-}_{4M_\pi^2}^{\Lambda^2} \frac{dx}{x^2(x-4M_\pi^2)}
\frac{I\!m\,f_1(x)}{x-s} \nonumber \\
&+&  \frac{1}{\pi} \int_{4M_\pi^2}^{\Lambda^2} \frac{dx}{x} \sum_{b=o}^2
W_{1b} (s,x) I\!m\,f_b(x) + O(p^8)
\end{eqnarray}
\end{mathletters}
Here, $P_a(s)$ are third order polynomials whose coefficients are
defined in terms of the six constants $t_0, t_2, t_3$ and $v_1, v_2,
v_3$ which appear in Eqs.~(3.6).  These polynomials are tabulated in
Appendix C, together with the nine kernels $W_{ab}(s,x)$ which define
the left-hand cut contributions to the partial waves.

Eqs.~(3.7a) and (3.7b) may be viewed as a particular truncation of the
infinite system of Roy equations, which slightly differs from the form
in which these equations have been used in the past \cite{bfp}.  Here,
the truncation in angular momentum and energy is performed under the
systematic control of chiral power counting.  In particular,
Eqs.~(3.7a,b) do not require a model-dependent evaluation of
``driving-terms'' which in the standard treatment behave in the chiral
limit as $O(p^4)$, owing to the use of twice-subtracted dispersion
relations. The price to pay is the occurrence of six ({\it a priori}
unknown) constants in the polynomials $P_a(s)$ instead of only two
constants (usually, the two $s$ wave scattering lengths) which
characterize the inhomogeneous terms in standard Roy equations
\cite{roy,bfp}.

Eqs.~(3.7a) and (3.7b) can be used to fully reconstruct from the data
the {\it whole amplitude $A(s|tu)$ up to and including accuracy
$O(p^6)$} in the whole low-energy domain of the Mandelstam plane,
including the unphysical region.  For this purpose one has to know the
absorptive parts of three lowest partial waves for $4M_\pi^2 < s <
\Lambda^2$ {\it and} the six constants $t_0, t_2, t_3,v_2,v_2,v_3$.
Suppose one knew $I\!m\,f_a(s)$ with associated error bars in the
whole interval $4M_\pi^2 < s < \Lambda^2 \lesssim 1~GeV^2$.  Then one
could calculate the dispersion integrals on the right hand side of
Eqs.~(3.7a,b).  One would then determine the constants $t$ and $v$ from
the best fit to the values $Re\,f_a(s)$ determined from the input
$I\!m\,f_a(s)$ via the unitarity condition.  The $\chi^2$ of this fit
may be considered as a measure of the internal consistency of the input
data $I\!m\,f_a(s)$.  In practice, {\it experimental} information on
$I\!m\,f_a(s)$ is only available for $s$ well above the threshold. In
this case, a more sophisticated iteration procedure \cite{iterate}of
Eqs.~(3.7a,b)
has to be used in order (i) to extrapolate the experimental data down
to the threshold and, simultaneously, (ii) to determine the six
constants $t$ and $v$. In both cases, the resulting amplitude is given
by the formula (3.2).

\subsection{\bf Proof of the reconstruction theorem}

Formulae (3.2) and (3.6) can be proven following the original
derivation of the Roy equations \cite{roy}.  The proof is based on
fixed $t$ dispersion relations for the three $s$-channel isospin
amplitudes $F^{(I)}$
\begin{equation}
{\bf F}(s,t,u) = \left ( \begin{array}{c} F^{(0)}\\F^{(1)}\\F^{(2)}\\
\end{array} \right ) (s,t,u),
\end{equation}
combined with the crossing symmetry relations
\begin{equation}
{\bf F}(s,t,u) = C_{su}{\bf F}(u,t,s) = C_{st}{\bf F}(t,s,u) =
C_{ut}{\bf F}(s,u,t).
\end{equation}
(Properties of the crossing matrices $C_{su},C_{st}$ and $C_{ut}$ are
reviewed in Appendix A.)  The standard Roy equations are derived from
twice-subtracted dispersion relations -- cf. the minimal number of
subtractions required by the Froissart bound.  In this case, however,
the high-energy tail of the dispersion integral, which is hard to
control in a model independent way, contributes to the $O(p^4)$ part
of the amplitude.  (In standard Roy equations, this contribution is
contained in the so-called driving terms \cite{bfp}.)  If, on the
other hand, one requires at low energy the precision $O(p^4)$ or
higher, then it is more appropriate to stick to less predictive {\it
triply-subtracted} fixed-$t$ dispersion relations:
\begin{eqnarray}
{\bf F}(s,t) &=& C_{st} \{ {\bf a_+}(t) + (s-u){\bf b_-}(t) + (s-u)^2
{\bf c_+}(t)  \} \nonumber \\
&+ &\frac{1}{\pi}\int_{4M_\pi^2}^{\infty} \frac{dx}{x^3} \left \{
\frac{s^3}{x-s} + \frac{u^3}{x-u}C_{su} \right \} I\!m\,{\bf F}(x,t).
\end{eqnarray}
Here the subscript $\pm$ refers to the eigenvalues $\pm 1$ of the
crossing matrix $C_{tu}$.  [Notice that in the $s$-channel isospin
basis (3.8), $C_{tu} = diag(+1,-1,+1)$.]  The subtraction term then
represents the most general quadratic function in $s$ (for fixed $t$)
symmetric under $s-u$ crossing.  By construction, the dispersion
integral in Eq.~(3.10) exhibits $s-u$ crossing symmetry too.  The task
is now to impose the remaining two crossing relations and to determine
the subtraction functions {\bf a},{\bf b}, and {\bf c}.  This can be
achieved, neglecting in Eq.~(3.10) contributions of chiral order
$O(p^8)$ and higher.

Let $\Lambda$ be a scale set by the threshold of production of
non-Goldstone particles.  Let us split the dispersion integral in
Eq.~(3.10) into low energy ($x \leq \Lambda^2$) and high energy ($x >
\Lambda^2$) parts.  For $4M_\pi^2 < s < \Lambda^2$, the imaginary part
can be written as
\begin{equation}
I\!m\, {\bf F} = I\!m\, \mbox{\boldmath $\Phi_+$}(s) + \left ( 1 +
\frac{2t}{s-4M_\pi^2} \right ) I\!m\, \mbox{\boldmath $\Phi_-$}(s) + {\bf
A}_{\ell \geq 2} (s,t),
\end{equation}
where the first two terms stand for the contributions of $s$ and $p$
waves:
\begin{equation}
I\!m\, \mbox{\boldmath $\Phi_+$}(s) = \left ( \begin{array}{c} I\!m\,
f_0^0 (s) \\
0 \\ I\!m\, f_0^2 (s) \end{array} \right ),~~~ I\!m\, \mbox{\boldmath
$\Phi_-$}(s) = \left ( \begin{array}{c} 0 \\ 3I\!m\, f_1^1 (s) \\ 0
\end{array}
\right ).
\end{equation}
${\bf A}_{\ell \geq 2}$ then collects the absorptive parts of all
higher partial waves.  The reason for this particular splitting
resides in the chiral counting mentioned at the beginning of this
section:  The first two terms in Eq.~(3.11) dominantly behave as
$O(p^4)$, whereas ${\bf A}_{\ell \geq 2}$ is suppressed to $O(p^8)$.
The dispersion integral ${\bf I}(s,t)$ in Eq.~(3.10) then splits into
three parts,
\begin{equation}
{\bf I}(s,t) = {\bf I}_{\ell < 2} (s,t) + {\bf I}_{\ell \geq 2} (s,t)
+ {\bf I}_H (s,t).
\end{equation}
${\bf I}_{\ell < 2} $ (${\bf I}_{\ell \geq 2} $) is the
contribution of low-energy $\ell < 2$ ($\ell \geq 2$) partial waves,
and ${\bf I}_H$ represents the high frequency part in
which no partial wave decomposition is performed.  Extracting from
${\bf I}_H$ its leading low energy behavior, one can write
\begin{equation}
{\bf I}_H (s,t) = (s^3 + u^3 C_{su}) {\bf H}_{\Lambda} + {\bf R}_H,
\end{equation}
where ${\bf H}_{\Lambda}$ are constants which can be expressed as
integrals over high-energy $\pi - \pi$ total cross sections, and the
remainder behaves at low energies as
\begin{equation}
{\bf R}_H = O([p/\Lambda]^8).
\end{equation}
The low-energy high angular momentum part ${\bf I}_{\ell \geq 2}$ is
also suppressed to $O(p^8)$, reflecting the leading behavior of the
absorptive part ${\bf A}_{\ell \geq 2}$ in the chiral limit and the
fact that the corresponding dispersion integral (3.10) extends over a
finite interval $x \in [4M_\pi^2,\Lambda^2]$.  Hence, it remains to
concentrate on the low-energy low angular momentum part ${\bf I}_{\ell
< 2}$.  Using Eq.~(3.11), one easily checks the identity
\begin{eqnarray}
{\bf I}_{\ell < 2} &=& \mbox{\boldmath $\Phi$}(s,t,u) - C_{st} \left
\{ \mbox{\boldmath $\Phi$}_+ (t) + \frac{s-u}{t-4M_\pi^2}
\mbox{\boldmath $\Phi$}_- (t) \right \} \nonumber \\
& &+ (4M_\pi^2 - 2t)(s^2 + u^2
C_{su})\frac{1}{\pi}\int_{M_\pi^2}^{\Lambda^2} \frac{dx}{x^3}
\frac{I\!m\, \mbox{\boldmath $\Phi$}_- (x)}{x-4M_\pi^2},
\end{eqnarray}
where
\begin{eqnarray}
\mbox{\boldmath $\Phi$}(s,t,u) &=& \left \{ \mbox{\boldmath $\Phi$}_+
(s) + \frac{t-u}{s-4M_\pi^2} \mbox{\boldmath $\Phi$}_- (s) \right \}
\nonumber \\
& & +C_{su} \left \{ \mbox{\boldmath $\Phi$}_+ (u)
+\frac{t-s}{u-4M_\pi^2} \mbox{\boldmath $\Phi$}_- (u) \right \}
\nonumber \\
& & +C_{st} \left \{ \mbox{\boldmath $\Phi$}_+ (t) +
\frac{s-u}{t-4M_\pi^2} \mbox{\boldmath $\Phi$}_- (t) \right \}
\end{eqnarray}
and \mbox{\boldmath $\Phi_\pm$} denote the following dispersion
integrals over the imaginary parts of low-energy $s$ and $p$ waves
[cf. Eq.~(3.12)]:
\begin{eqnarray}
\mbox{\boldmath $\Phi$}_+ (s) &=& \frac{s^3}{\pi}
\int_{4M_\pi^2}^{\Lambda^2} \frac{dx}{x^3} \frac{I\!m\,
\mbox{\boldmath $\Phi$}_+ (x)}{x-s} \nonumber \\
\mbox{\boldmath $\Phi$}_- (s) &=& \frac{s^2(s-4M_\pi^2)}{\pi}
\int_{4M_\pi^2}^{\Lambda^2} \frac{dx}{x^2(x-4M_\pi^2)} \frac{I\!m\,
\mbox{\boldmath $\Phi$}_- (x)}{x-s}.
\end{eqnarray}
One observes from Eq.~(3.17) that the function \mbox{\boldmath
$\Phi$}$(s,t,u)$ exhibits the full three-channel crossing symmetry.
Furthermore, the second and third terms in Eq.~(3.16) represent a
function that is quadratic in $s$ (at fixed $t$) and symmetric under
$s-u$ crossing. These terms can therefore be absorbed into the
subtraction polynomial in the dispersion relations (3.10) by a
suitable redefinition of (yet unknown) subtraction functions ${\bf
a}_+,{\bf b}_-,{\bf c}_+$. Consequently, the whole amplitude ${\bf F}$
can be rewritten as
\begin{equation}
{\bf F}(s,t) = \mbox{\boldmath $\Phi$}(s,t,u) + {\bf P}(s,t,u) +
O([p/\Lambda]^8),
\end{equation}
where {\bf P} is of the form
\begin{eqnarray}
{\bf P} =& C_{st} \left \{ \mbox{\boldmath $\alpha$}_+ (t) + (s-u)
\mbox{\boldmath $\beta$}_- (t) + (s-u)^2 \mbox{\boldmath $\gamma$}_+
(t) \right \} \nonumber \\
&+ (s^3 + u^3 C_{su}) {\bf H}_{\Lambda}.
\end{eqnarray}
Notice that the unspecified $O(p^8)$ contributions in Eq.~(3.19)
originate both from the high-energy remainder ${\bf R}_H$ (3.14)
and from the low-energy higher angular momentum part ${\bf I}_{\ell
\geq 2}$.
Crossing symmetry of the scattering amplitude {\bf F} should hold
order by order in the chiral expansion.  Since the function
\mbox{\boldmath $\Phi$} (3.17) exhibits full crossing symmetry, it
remains to impose the latter for the function {\bf P} (3.20).  Because
of the manifest $s-u$ symmetry, it is enough to require
\begin{equation}
{\bf P}(s,t,u) = C_{st}{\bf P}(t,s,u).
\end{equation}
Neglecting $O(p^8)$ contributions, this equation represents the
necessary and sufficient condition for the complete crossing symmetry
of the amplitude {\bf F}.

Eq.~(3.21) can be easily solved.  Considering $s$ and $t$ as
independent variables, one easily finds that \mbox{\boldmath
$\alpha_+$}$(t)$, \mbox{\boldmath $\beta_-$}$(t)$, and \mbox{\boldmath
$\gamma_+$}$(t)$ should be cubic, quadratic and linear functions of
$t$ respectively.  Hence, {\bf P}$(s,t,u)$ is a general crossing
symmetric polynomial in the Mandelstam variables of (at most) third
order.  Such a polynomial contains six independent parameters (see
Appendix A).  Indeed, after some simple but lengthy algebra, one
verifies that Eq.~(3.21) leaves a six parameter freedom in the original
expression (3.20) for {\bf P}.

It remains to rewrite the result (3.19) in terms of the single
amplitude
\begin{equation}
A(s|tu) = A(s|ut) = \frac{32\pi}{3} \left \{ F^{(0)}(s,t,u) -
F^{(2)}(s,t,u) \right \}.
\end{equation}
The function \mbox{\boldmath $\Phi$} gives rise to a contribution of
the form (3.2) in which only the dispersion integrals of Eq.~(3.6)
occur.  (One easily checks that $I\!m\, T,~I\!m\, U$ and $I\!m\, V$
are given by Eqs.~(3.4).)  Furthermore, taking into account the
ambiguity (3.5) in the definition of $T,~U$ and $V$, it is clear that a
general crossing symmetric polynomial may be conveniently parametrized
by the six independent parameters $t_0,t_2,t_3,v_1,v_2,v_3$ as in
Eqs.~(3.6).

\section{\bf Perturbative $\pi-\pi$ amplitude and the \protect \\
effective infrared dimension of the quark mass}

We are now in a position to compare the two alternative low-energy
expansions of the amplitude $A(s|tu)$ generated by chiral perturbation
theory according to the two possible values of the effective dimension of
the quark mass: 2, in the case of the standard $\chi$PT, and 1 in the case
which was defined in Sec. II as improved $\chi$PT.   Up to and including
two loops, the amplitude $A$ should be of the general form (3.2).
Consequently, neglecting $O(p^8)$ contributions, one can work with the
three functions $T, U$ and $V$ of a single variable and decompose them
as
\begin{equation}
 T(s) = \sum_{n=0}^2 T^{(n)}(s),~ U(s) = \sum_{n=0}^2
   U^{(n)}(s),~ V(s) = \sum_{n=0}^2 V^{(n)}(s),
\end{equation}
where $n$ refers to the number of loops (including tree contributions
of the corresponding order).  It will be shown that the amplitudes $T,
U$ and $V$ start to be sensitive to the effective dimension of the
quark mass at leading $(n=0)$, one-loop $(n=1)$ and two-loop levels
respectively.

\subsection{\bf Leading $O(p^2)$ order}

If the dimension of the quark mass is 2, {\it i.e.}, if each power of
the scalar pseudoscalar source $\chi$ in ${\cal L}_{eff}$ counts for
two powers of pion momentum (mass), then the effective Lagrangian is
dominated by the well-known expression
\begin{equation}
{\cal L}^{(2)} = \frac{1}{4} F_0^2
    \{\langle (D^\mu U)^+ (D_\mu U) \rangle + 2B_0
    \langle \chi^+ U + U^+ \chi \rangle \}.
\end{equation}
This formula collects all possible invariants of dimension 2. To
leading order, the pion and the kaon masses read
\begin{eqnarray}
\overcirc{M}_\pi^2 &=& 2\hat{m}B_0 \nonumber \\
\overcirc{M}_K^2 &=& (m_s + \hat{m}) B_0
\end{eqnarray}
and the $\pi-\pi$ amplitude takes the well-known form, first given by
Weinberg \cite{sw66}
\begin{equation}
A_{lead} (s|tu) = \frac{1}{F_0^2} (s- 2\hat{m}B_0) = \frac{1}{F_0^2}
(s - \overcirc{M}_\pi^2)
\end{equation}
This represents the standard scenario of chiral perturbation theory.
It can hardly be circumvented provided the scale of $B_0$ is large
compared to the pion mass, typically, $B_0 \gtrsim 1~GeV$.

On the other hand, if $B_0$ turned out to be much smaller than the
GeV-scale, {\it e.g.}, comparable to the fundamental order parameter
$F_0$ ($\sim 93~ MeV$), then the above way of counting effective
infrared dimensions would be modified.  Both the quark mass
and the condensate $B_0$ should then be considered as quantities
comparable to the pion mass.  They should both be attributed effective
infrared dimension 1 and they should both be viewed as
expansion parameters.  In this case, every insertion of the
source $\chi(x)$ counts as a single power of pion momentum
and the formula (4.2) no longer represents the most general
expression of dimension 2.  Instead, the complete collection of
invariants of dimension 2 now reads
\begin{eqnarray}
    \tilde{{\cal L}}^{(2)} &=& \frac{1}{4} F_0^2 \{ \langle D^\mu
   U D_\mu U^+ \rangle + 2B_0 \langle \chi^+ U +
   \chi U^+\rangle \nonumber \\
     &+& A_0 \langle \chi^+ U \chi^+ U + \chi U^+ \chi U^+ \rangle
     + Z_0^S \langle \chi^+ U + U^+ \chi \rangle^2 + \nonumber \\
     &+& Z_0^P \langle \chi^+ U - \chi U^+ \rangle ^2 +
     2H_0 \langle \chi^+ \chi \rangle \} .
\end{eqnarray}
where the tilde over the symbol ${\cal L}$ here (and below) indicates
the use of the modified chiral power counting. The terms containing
two powers of $\chi$ are usually included into the next-to-the-leading
part ${\cal L}^{(4)}$ of the effective Lagrangian. Here, they appear
of the same dimension and they are expected to be of a comparable size
as the standard expression (4.2). The low-energy constants $A_0,
Z_0^S$ and $Z_0^P$ represent appropriately subtracted zero-momentum
transfer two-point functions of scalar and pseudoscalar quark
densities, divided by $F_0^2$.  These two-point functions are order
parameters of spontaneous chiral symmetry breaking and, consequently,
they satisfy superconvergent dispersion relations.  A simple
saturation of the latter with a few of the lowest massive hadronic
states suggests that the dimensionless constants $A_0$ and $Z_0^P$ are
of the order $1$, say, $A_0 \sim 1 - 5$.  On the other hand, $Z_0^S$
violates the Zweig rule in the $0^{++}$ channel and consequently it is
expected to be suppressed.  The parameters $Z_0^S, Z_0^P$ and $A_0$
are related to the low-energy constants $L_6, L_7$ and $L_8$ of the
standard $d = 4$ Lagrangian ${\cal L}^{(4)}$ \cite{gl85}.  Expanding
the latter constants in powers of $B_0$, one gets\footnote{The order
of magnitude estimate $A_0 \sim 1 - 5$ is compatible with the standard
$\chi$PT estimates.  Taking $A_0 \sim 5$, and using the standard value
$B_0 \sim 1.2~GeV$, the $A_0$-contribution to $L_8$ in Eq.~(4.6)
becomes $1.6 \times 10^{-3}$, which is consistent with the standard
$\chi$PT measurement of $L_8$ \cite{gl85}.\label{A_0}}
\begin{eqnarray}
    L_6 &=& \left ( \frac{F_0}{4B_0} \right ) ^2 \{ Z_0^S +
O(B_0^2)\},\nonumber \\
    L_7 &=& \left ( \frac{F_0}{4B_0} \right ) ^2 Z_0^P \nonumber \\
    L_8 &=& \left ( \frac{F_0}{4B_0} \right ) ^2 \{A_0 + O(B_0^2) \},
\end{eqnarray}
where $O(B_0^2)$ terms represent divergent contributions to the
two-point functions defining the divergent parts of the bare constants
$L_6, L_8$.  (The constants $A_0, Z_0^S$ and $Z_0^P$ do not undergo
any infinite renormalization.)

The leading order pion and kaon masses (denoted by a tilde) now read
\begin{eqnarray}
\widetilde{M}_\pi^2 &=& 2\hat{m}(\tilde{B} + 4 \hat{m} Z_0^S)
    + 4\hat{m}^2 A_0 \nonumber \\
\widetilde{M}_K^2 &=& (m_s + \hat{m}) (\tilde{B} + 4 \hat{m} Z_0^S)
    + (m_s + \hat{m})^2 A_0 .
\end{eqnarray}
Here $\tilde{B}$ stands for the dominant $O(p)$ contribution to the
$SU(2) \times SU(2)$ quark-antiquark condensate (divided by $F_0^2$)
taken at $m_u = m_d = 0$:
\begin{equation}
\langle \bar{u}u \rangle_{m_u=m_d=0} = \langle \bar{d}d
\rangle_{m_u=m_d=0} = -F_0^2 \tilde{B} + O(m_s^2).
\end{equation}
Within the modified chiral power counting, $\tilde{B}$ consists of two
terms
\begin{equation}
\tilde{B} = B_0 + 2 m_s Z_0^S
\end{equation}
which are both of the order $O(p)$.  In principle, they could be of
comparable size, if $Z_0^S$ were not suppressed by the Zweig rule.

The leading contribution to the $\pi-\pi$ scattering amplitude
calculated from the improved $O(p^2)$ Lagrangian (4.5) turns out to be
independent of low-energy parameters $A_0$ and $Z_0^P$, and it can be
expressed in term of the quark-antiquark condensate $\tilde{B}$,
\begin{equation}
  A_{lead} (s|tu) = \frac{1}{F_0^2} (s - 2\hat{m} \tilde{B}),
\end{equation}
in complete analogy with the standard result (4.4).  Although
Eq.~(4.10) and Weinberg's formula (4.4) formally coincide if one
neglects Zweig-rule violation, their numerical content is rather
different, because of different scales of quark-antiquark condensation
in each $\chi$PT alternative.  In Eq.~(4.4), $2\hat{m}B_0$ is the leading
approximation to $M_\pi^2$, whereas in the improved $\chi$PT, the relation
between the quark-antiquark condensate and the pion mass is more
subtle:  Indeed, using first Eq.~(4.7), formula (4.10) can be rewritten
as
\begin{equation}
A_{lead} (s|tu) = \frac{1}{F_0^2} (s - \widetilde{M}_\pi^2) +
\frac{\widetilde{M}_\pi^2}{F_0^2}\,\epsilon \,(1 + 2\zeta),
\end{equation}
where
\begin{equation}
  \epsilon = \frac{4\hat{m}^2 A_0}{\widetilde{M}_\pi^2},
     ~~~~~~~~~~~ \zeta = \frac{Z_0^S}{A_0}.
\end{equation}
Whereas in the standard $\chi$PT $\epsilon$ would be a small quantity of
the order $O(p^2)$, in the improved $\chi$PT, $\epsilon$ is $O(1)$ and
there is no reason for it to be particularly small; hence, the second
term in Eq.~(4.11) represents a {\it leading order} modification of
the Weinberg's formula (4.4). ($\zeta$ measures the Zweig rule
violation in the $0^{++}$ channel and can be expected rather small.)
Using Eqs.~(4.7) one may easily check that $\epsilon$ can indeed be of
order 1 for natural values of $A_0$ (cf. footnote \ref{A_0}) and for
reasonably small values of quark masses. Setting --- for the sake of
illustration --- $B_0 = 150~MeV$ and $\hat{m} = 25~MeV$, and
neglecting Zweig rule violation, one obtains $\epsilon = 0.62, A_0 =
4.8$ and $m_s \simeq 195~MeV$.

The leading order mass formula (4.7) implies a relation between the
parameter $\epsilon$ and the quark-mass ratio $ r = m_s/\hat{m}$:
\begin{equation}
  \epsilon = 2 \frac{r_2-r}{r^2-1} ,~~~ r_2 =
2\frac{\widetilde{M}_K^2}{\widetilde{M}_\pi^2} - 1 \simeq 25.9
\end{equation}
If $r$ decreases from its canonical leading order value $r = r_2$,
then $\epsilon$ increases and reaches 1 for $r = r_1$,
\begin{equation}
 r_1 = 2 \frac{\widetilde{M}_K}{\widetilde{M}_\pi} - 1 \simeq 6.33.
\end{equation}
Similarly, the order parameter $B_0$ can be expressed as
\begin{equation}
\frac{2\hat{m}B_0}{\widetilde{M}_\pi^2} = 1 - [1 +
(r+2)\zeta]\epsilon.
\end{equation}
This ratio decreases from its canonical value 1 down to zero, as $r$
decreases from $r=r_2$ to $r=r_{crit}(\zeta) \gtrsim r_1$, for which
$B_0$ vanishes.  Notice that stability of the massless QCD vacuum
under perturbation by small quark masses implies $B_0 \geq 0$.

\subsection{\bf Next to the leading $O(p^3)$ contribution}

In the improved chiral perturbation theory, the leading order
Lagrangian $\tilde{{\cal L}}^{(2)}$ is followed by a {\it dimension 3
term} $\tilde{{\cal L}}^{(3)}$, which contributes at the tree level
before one-loop contributions of dimension 4 start to appear.
$\tilde{{\cal L}}^{(3)}$ reads
\begin{eqnarray}
    \tilde{{\cal L}}^{(3)} &=& \frac{1}{4} F_0^2 \{ \xi \langle D_\mu U^+
      D^\mu \chi + D_\mu \chi^+ D^\mu U \rangle \nonumber \\
    &+& \rho_1 \langle (\chi^+U)^3 + (\chi U^+)^3 \rangle
     + \rho_2 \langle \chi ^+ \chi (\chi^+ U + U^+ \chi) \rangle
\nonumber \\
    &+& \rho_3 \langle (\chi^+ U)^2 - (\chi U^+)^2 \rangle
       \langle \chi^+ U - \chi U^+ \rangle + \ldots \} .
\end{eqnarray}
The dots stand for terms that violate the Zweig rule in a nonanomalous
channel.  Notice that (4.16) differs in its first term from the
expression given for $\tilde{{\cal L}}^{(3)}$ in Ref.\ \cite{fss91}.
The two forms of $\tilde{{\cal L}}^{(3)}$ are equivalent: they are
related by a simple redefinition of the Goldstone boson field $U$.
The low energy constants $\xi$ and $\rho_i$ are finite --- there are
no divergences of dimension 3.  $\tilde{{\cal L}}^{(3)}$ induces a
shift in the pion mass,
\begin{equation}
\delta M_\pi^2 = \epsilon \, \widetilde{M}_\pi^2 \, (9\lambda_1 +
\lambda_2),
\end{equation}
where
\begin{equation}
  \lambda_i = \frac{\hat{m}\rho_i}{4A_0}
\end{equation}
are dimensionless parameters of order $O(M_\pi)$.  Similarly,
the leading $\pi-\pi$ amplitude receives a constant $d = 3$
contribution
\begin{equation}
 \delta \tilde{A} (s|tu) = \epsilon \,
\frac{\widetilde{M}_\pi^2}{3F_0^2} \, (81 \,\lambda_1 + \lambda_2).
\end{equation}
Finally, the first term in $\tilde{{\cal L}}^{(3)}$ is responsible for
splitting of the decay constants $F_\pi, F_K, F_\eta$.  Eliminating
the low-energy parameter $\xi$, one obtains, to that order
\begin{equation}
  \frac{F_\pi^2}{F_0^2} = 1 + \frac{2}{r-1}
     (\frac{F_K^2}{F_\pi^2} - 1).
\end{equation}

It is convenient to collect all $d=2$ and $d=3$ contributions, and to
express the resulting tree amplitude in the form (3.1):
\begin{equation}
 A_{tree} (s|tu) = \frac{1}{3F_\pi^2} [ \alpha M_\pi^2 + \beta
(3s-4M_\pi^2)] + \frac{M_\pi^2}{3F_\pi^2} \,\delta \alpha ,
\end{equation}
where $M_\pi$ and $F_\pi$ denote the $experimental$ (charged)
pion mass and decay constant.\footnote{In practice, $M_\pi$ = 139.6
MeV and $F_\pi$ = 93.1 MeV will be identified with the corresponding
theoretical expressions up to and including the highest order of $\chi$PT
considered.} The parameters $\alpha$ and $\beta$ read
\begin{equation}
\frac{\alpha}{\beta} = 1 + 3 \epsilon \, (1+2\zeta),~~~~\beta =
\frac{F_\pi^2}{F_0^2},
\end{equation}
whereas $\delta \alpha = \delta \alpha_3 + \delta \alpha_4$ describes
small $O(p^3)$ and $O(p^4)$ corrections.  $\delta \alpha$ arises from
the genuine $O(p^3)$ contribution (4.19) of $\tilde{\cal L}^{(3)}$ to
the $\pi-\pi$ amplitude and from the introduction of the physical mass
$M_\pi$ into the formula (4.21).  Using Eqs.~(4.17) and (4.19), the
$O(p^3)$ constant $M_\pi^2~\delta \alpha_3$ can be expressed in terms
of the parameters $\lambda_1$ and $\lambda_2$ of $\tilde{\cal
L}^{(3)}$:
\begin{equation}
M_\pi^2 ~ \delta \alpha_3 = \epsilon \, \beta \,
\widetilde{M}_\pi^2 \, [72 \lambda_1 - (27\lambda_1 +
3 \lambda_2)\, \epsilon \, (1+2\zeta)].
\end{equation}
The remaining term $M_\pi^2 \, \delta \alpha_4$ accounts for the
$O(p^4)$ and higher contributions to $M_\pi^2$.  One has
\begin{equation}
M_\pi^2 \, \delta \alpha_4 = - \beta \, \Delta \! M_\pi^2 \, [1 +
3\epsilon(1+2\zeta)],
\end{equation}
where
\begin{equation}
\Delta \! M_\pi^2 = M_\pi^2 - \widetilde{M}_\pi^2 - \delta M_\pi^2
\end{equation}
represents the $O(p^4)$ difference between the physical value and the
tree approximation of the pion mass squared.

The results of standard $\chi$PT are reproduced by setting $\epsilon =
\zeta = 0$ in the previous equations; $i.e.$, $r = r_2 \simeq 25.9$.
In this case, $\widetilde{M}_\pi^2$ reduces to $\overcirc{M}_\pi^2$
[Eq.~(4.3)], and $\alpha = \beta \simeq 1$.  The improved $\chi$PT still
requires $\beta \simeq 1$, but $\alpha$ is now allowed and expected to
be considerably larger, since $\epsilon$ is now an $O(1)$ quantity.
In fact, the vacuum stability conditions mentioned above imply that
for a given quark mass ratio $r$ [lying between $r_1$ and $r_2$ -- cf.
Eqs.~(4.13) and (4.14)], the Zweig rule violating parameter $\zeta =
Z_0^S/A_0$ should satisfy
\begin{equation}
0 \leq \zeta \leq \zeta_{crit}(r) = \frac{1}{2} \frac{r-r_1}{r_2 - r}
\frac{r+r_1+2}{r+2} .
\end{equation}
Using these bounds in Eq.~(4.22), one obtains a rather narrow band of
allowed values in the plane defined by the ratio $\alpha/\beta$ and
$r$.  This band is shown in Fig. 1.

It is straightforward to rewrite the above result in terms of the
amplitudes $T,U$ and $V$.  The tree contribution to these amplitudes
simply reads
\begin{equation}
  T^{(0)} (s) = (\hat{\alpha} + \delta \hat{\alpha}) M_\pi^2,~~
U^{(0)}(s) = 0,~~     V^{(0)}(s) = 9\hat{\beta},
\end{equation}
where
\begin{equation}
  \hat{\alpha} \equiv \frac{\alpha}{96\pi}\frac{1}{F_\pi^2},
{}~~~~~~~~~~~~ \hat{\beta} \equiv \frac{\beta}{96\pi}\frac{1}{F_\pi^2}
\end{equation}
and likewise for $\delta \hat{\alpha}$.  Our main task is to use all
available experimental information to measure $\alpha,\beta$ and,
indirectly, the quark mass ratio $r$.

\subsection{\bf One loop $O(p^4)$ order}

Let ${\cal L}_{nm}$ denote an invariant entering the effective
Lagrangian, that contains $n$ powers of covariant derivatives D and
$m$ insertions of the scalar-pseudoscalar source $\chi$. (For
simplicity, the expansion coefficients $\ell^{nm}$ of Eq.\ (2.7) are
included in ${\cal L}_{nm}$.)  In the standard chiral perturbation
theory the $d=4$ part of the effective Lagrangian can be written as
\begin{equation}
  {\cal L}^{(4)} = \sum_{n+2m=4} {\cal L}_{nm}.
\end{equation}
It contains all counterterms which are needed to renormalize one-loop
contributions generated by ${\cal L}^{(2)}$.  If the dimension of the
quark mass is 1, one-loop renormalization gets modified in two
respects: (i) The effective dimension of a term ${\cal L}_{nm}$ is $d
= n+m$ instead of $d = n+2m$, and (ii) $B_0$ is now a (small)
expansion parameter of dimension 1.  It follows, in particular, that
renormalization has to be performed order by order in $B_0$.  The
modified $d=4$ part of ${\cal L}_{eff}$ then reads
\begin{equation}
  \tilde{{\cal L}}^{(4)} = \sum_{n+m=4} {\cal L}_{nm} + B_0 \,
     ({\cal L}_{21} + {\cal L}_{03}) + B_0^2 \, {\cal L}_{02}.
\end{equation}
The last two counterterms are needed to renormalize the
$B_0$-dependent part of one-loop divergences generated by
$\tilde{{\cal L}}^{(2)}$.  Terms which are contained both in ${\cal
L}^{(4)}$ and in $\tilde{{\cal L}}^{(4)}$ are merely made with four
derivatives \cite{gl85}:
\begin{eqnarray}
    {\cal L}_{40} &=& L_1 \langle D_\mu U^+ D^\mu U\rangle^2
       + L_2 \langle D_\mu U^+ D_\nu U\rangle \langle D^\mu U^+
       D^\nu U\rangle \nonumber \\
    &+& L_3 \langle D_\mu U^+ D^\mu U D_\nu U^+ D^\nu U \rangle
\nonumber \\
    &-& iL_9 \langle F_{\mu\nu}^R D^\mu UD^\nu U^+ +
      F_{\mu\nu}^L D^\mu U^+ D^\nu U\rangle \nonumber \\
    &+& L_{10} \langle U^+ F_{\mu\nu} ^R U F^{L,\mu\nu}\rangle
      + H_1 \langle F_{\mu\nu}^R F^{R,\mu\nu} + F_{\mu\nu}^L
         F^{L,\mu\nu}\rangle .
\end{eqnarray}
The meaning and renormalization of low-energy constants in Eq.~(4.31)
are independent of the symmetry breaking sector and, in particular, of
the infrared dimension of the quark mass.  The remaining
$B_0$-independent terms in Eq.~(4.30), cf.  ${\cal L}_{22}$ and ${\cal
L}_{04}$, are absent from the expression for ${\cal L}^{(4)}$: with
quark mass of dimension 2, these terms would count as $O(p^6)$ and
$O(p^8)$ respectively.  On the other hand, all terms but ${\cal
L}_{40}$ contained in ${\cal L}^{(4)}$ are already included either in
$\tilde{{\cal L}}^{(2)}$ or in $\tilde{{\cal L}}^{(3)}$.
Consequently, $\tilde{{\cal L}}^{(2)} + \tilde{{\cal L}}^{(3)} +
\tilde{{\cal L}}^{(4)}$ not only encompasses all terms of the standard
${\cal L}^{(2)} + {\cal L}^{(4)}$ but, in addition, it contains new
terms of the type $\tilde{{\cal L}}^{(3)}, {\cal L}_{22}$ and ${\cal
L}_{04}$.  This phenomenon is general.  Order by order, the improved
$\chi$PT contains the standard perturbation theory as a special case:  It
contains more parameters and it could well fit the experimental data
even when the standard $\chi$PT fails.

The one-loop contribution to the $\pi-\pi$ amplitude $A(s|tu)$ has
been worked out within the standard chiral perturbation theory in Refs.
\cite{gl84,gl85}.  The result can be expressed in terms of four constants:
$\alpha =\beta$ (close to 1), the shift $\delta \alpha_4$ ($\delta
\alpha_3 = 0$ in this case) introduced in Eqs.~(4.21), and two linear
combinations of the renormalized constants $L_1, L_2$ and $L_3$.  In
the improved $\chi$PT, the one-loop $O(p^4)$ amplitude contains, in
addition, two parameters which arise from the new terms ${\cal
L}_{22}$ and ${\cal L}_{04}$ in $\tilde{{\cal L}}^{(4)}$.  Working
with the amplitudes $T,U,V$ (the formula (3.2) is valid up to and
including two loops), one may obtain a closed form for the one-loop
amplitude which encompasses both alternatives of chiral perturbation
theory.

Let $\varphi_a^{(d)}(s)$ denote the effective dimension-$d$
contribution to the real part of the partial wave amplitude
$f_a(s),~(a=0,1,2)$, introduced in section III B:
\begin{equation}
Re\,f_a(s) = \sum_{d \geq 2} \varphi_a^{(d)}.
\end{equation}
{}From Eqs.~(4.27) one finds
\begin{eqnarray}
    \varphi_0^{(2)}(s) &=& 6\hat{\beta}\, (s + \kappa_0) \nonumber \\
    \varphi_1^{(2)}(s) &=& \hat{\beta}\, (s-4M_\pi^2) \nonumber \\
    \varphi_2^{(2)}(s) &=& -3\hat{\beta}\,(s + \kappa_2),
\end{eqnarray}
where
\begin{equation}
  \kappa_0 \equiv (\frac{5\alpha}{6\beta} - \frac{4}{3})
     M_\pi^2,~~~~~ \kappa_2 \equiv (-\frac{2\alpha}{3\beta} - \frac{4}{3})
      M_\pi^2 .
\end{equation}
Similarly, the real parts at the $O(p^3)$ level are
\begin{equation}
\varphi_0^{(3)} = 5 M_\pi^2 \, \delta \alpha_3,~~~\varphi_1^{(3)} =
0,~~~ \varphi_2^{(3)} = -2 M_\pi^2 \, \delta \alpha_3,
\end{equation}
where $\delta \alpha_3$ is given by Eq.~(4.23).  For $d > 3$, the real
parts are no longer defined by the tree amplitude alone.  The $O(p^d)$
contribution to the imaginary part of the partial wave amplitudes
$I\!m\, f^{(d)}_a(s)$ can be expressed for $s > 4 M_\pi^2$ through elastic
unitarity:
\begin{equation}
  I\!m \, f^{(d)}_a(s) = \sqrt{\frac{s-4M_\pi^2}{s}} \sum_{d_1 + d_2 = d}
\varphi_a^{(d_1)}(s) \varphi_a^{(d_2)}(s).
\end{equation}
This result is an exact property of $\chi$PT amplitudes for $4 \leq d <
8$.

The one-loop level contains $d=4$, $d=5$ and $d=6$ contributions to
the scattering amplitude $A(s|tu)$.  In the following, we shall merely
concentrate on the leading $O(p^4)$ part.  The corresponding
components of the functions $T,U,V$\footnote{Notice that $O(p^N)$
terms in $V$ contribute to the scattering amplitude $A$ of Eq.~(3.2) as
$O(p^{N+2})$.} will be denoted as $T_{lead}^{(1)}(s),
U_{lead}^{(1)}(s)$, and $V_{lead}^{(1)}(s)$.  The discontinuities of
these functions are given by the $O(p^4)$ absorptive parts
$I\!m\,f_a^{(4)}$, following Eqs.~(3.4).  Hence, the $O(p^4)$ one-loop
amplitudes $T,U,V$ can be written as
\begin{eqnarray}
T_{lead}^{(1)}(s) &=& \frac{1}{3} \{ [\varphi^{(2)}_0(s)]^2 + 2
[\varphi^{(2)}_2 (s)]^2 \} L(s,\mu^2) + \alpha_4(\mu^2) +
\alpha_0(\mu^2)s^2 \nonumber \\
U_{lead}^{(1)}(s) &=& \frac{1}{2} \{2[\varphi^{(2)}_0(s)]^2 - 5
[\varphi^{(2)}_2(s)]^2\} L(s,\mu^2) \nonumber \\
V_{lead}^{(1)}(s) &=& \frac{27}{2}
\frac{[\varphi^{(2)}_1(s)]^2}{s-4M_\pi^2} L(s, \mu^2) + \beta_2(\mu^2)
+ \beta_0 (\mu^2)s,
\end{eqnarray}
where $L(s,\mu^2)$ is the loop integral subtracted at the point $s =
-\mu^2$:
\begin{equation}
L(s,\mu^2) \equiv \frac{s+\mu^2}{\pi} \int_{4M_\pi^2}^\infty
\frac{dx}{x+\mu^2} \frac{1}{x-s} \sqrt{\frac{x-4M_\pi^2}{x}}.
\end{equation}
The constants $\alpha_n(\mu^2)$ and $\beta_n(\mu^2)$ behave in the
chiral limit as $M_\pi^n$.  They describe the most general polynomial
part of $T,~U$ and $V$ which is $ O(p^4)$ and takes into account the
freedom (3.5). These constants represent renormalized tree
contributions of the $d=4$ part of ${\cal L}_{eff}$.  Their dependence
on the subtraction point $\mu^2$ can be determined by demanding that
the scattering amplitude $A(s|tu)$ be $\mu^2$-independent.  Following
Appendix B, this requirement is equivalent to the conditions
\begin{equation}
\frac{\partial}{\partial \mu^2} T^{(1)}(s) =
\delta T(s), ~~\frac{\partial}{\partial \mu^2} U^{(1)}(s) =
\delta U(s),~~\frac{\partial}{\partial \mu^2} V^{(1)}(s) = \delta V(s),
\end{equation}
where $\delta T,~ \delta U$ and $\delta V$ are of the general
form (3.5).  Taking into account the $s$-independence of
$\frac{\partial}{\partial \mu^2} L (s,\mu^2)$ and $L(0,\mu^2) =
-L(-\mu^2,0)$, the solution of Eqs.~(4.39) can be easily found:
\begin{eqnarray}
    \alpha_0(\mu^2) &=& \alpha_0 (0) + 18 \hat{\beta}^2 L(-\mu^2)
\nonumber \\
    \beta_0(\mu^2) &=& \beta_0 (0) \nonumber \\
    \alpha_4(\mu^2) &=& \alpha_4(0) + (11 \hat{\alpha}^2 - 32
      \hat{\beta}^2) M_\pi^4 L(-\mu^2) \nonumber \\
    \beta_2(\mu^2) &=& \beta_2(0) + 18 \hat{\beta} (5\hat{\alpha}
      - 2\hat{\beta}) M_\pi^2 L(-\mu^2).
\end{eqnarray}
In these equations we have denoted
\begin{equation}
  L(s) \equiv L(s,\mu^2 = 0) = \frac{1}{\pi} \left [ 2 + \sigma
    \ln (\frac{\sigma - 1}{\sigma +1}) \right ] ,~~~~
    \sigma = \sqrt{1 - \frac{4M_\pi^2}{s}}.
\end{equation}
In the following we shall work at $\mu^2 = 0$.

The constants $\alpha_0$ and $\beta_0$ are related to the low-energy
parameters $L_1, L_2$ and $L_3$ which occur in the expression (4.31)
for ${\cal L}_{40}$.  One gets
\begin{eqnarray}
   \alpha_0(0) &=& \frac{1}{4\pi F_0^4} [ L_1^r + L_2^r
      + \frac{1}{2} L_3 - \frac{1}{4}\nu (\bar{\mu}^2) ] \nonumber \\
   \beta_0(0) &=& \frac{3}{8\pi F_0^4} (L_2^r - 2L_1^r -L_3)
    + \frac{1}{1024\pi^3F_0^4}
\end{eqnarray}
where
\begin{equation}
\nu (\bar{\mu}^2) = \frac{1}{32\pi^2} \left [ \ln
    \frac{M_\pi^2}{\bar{\mu}^2} + \frac{1}{8} \ln
\frac{M_K^2}{\bar{\mu}^2} + \frac{9}{8} \right ]
\end{equation}
and $\bar{\mu}^2$ denotes the renormalization scale introduced in Ref.\
\cite{gl85}.  The renormalized constants $L_1^r, L_2^r$ are
$\bar{\mu}^2$-dependent, whereas $L_3$ and $L_2^r - 2\,L_1^r$ are not.
Furthermore, the combination $L_2^r - 2\,L_1^r$ should be suppressed
by the Zweig rule or in the large $N_c$ limit.  Notice that the
constant $\beta_0$ is independent both of $\mu^2$ and of
$\bar{\mu}^2$. The interpretation of the remaining two constants
$\alpha_4$ and $\beta_2$ depends on the effective dimension of the
quark mass. In the standard chiral perturbation theory, these
constants can be expressed in terms of the shifts of the pion mass and
decay constant, as calculated within $SU(2) \times SU(2)$ perturbation
theory \cite{gl84}. In the improved $\chi$PT, $\alpha_4(0)$ and
$\beta_2(0)$ are independent parameters which describe respective
contributions of new terms ${\cal L}_{04}$ and ${\cal L}_{22}$ in the
$O(p^4)$ effective Lagrangian $\tilde{\cal L}^{(4)}$.  The explicit
relationship between $\alpha_4(0), \beta_2(0)$ and the low-energy
parameters of $\tilde{\cal L}^{(4)}$ is of no direct use in the
present paper and it will be given elsewhere.

Concluding this section, it is worth noting that the low-energy
theorem of Sec. III considerably simplifies the calculation of
two-loop contributions to $A(s|tu)$:  For $d < 8$, all $O(p^d)$ terms
can be obtained by a straightforward combination of Eqs.~(3.2) and
(3.4) with the unitarity condition (4.36).  Up to and including two
loops, the $\chi$PT expansion of the $\pi-\pi$ scattering amplitude can be
viewed as an iteration of the Roy-type Eqs.~(3.7a) and (3.7b).  The
corresponding polynomials $P_a(s)$ appearing at a given order $O(p^d)$
are then defined in terms of the renormalized low-energy constants of
the Lagrangians ${\cal L}^{(d)}$ or $\tilde{\cal L}^{(d)}$, according
to the effective dimension of the quark mass being respectively 2 or
1.
\section{\bf Determination of parameters of ${\cal L}_{\lowercase
{eff}}$ \protect \\ from $\pi-\pi$ scattering data}

Suppose one has enough experimental information to perform the program
formulated in Sec. III and to reconstruct the low-energy amplitude
$A(s|tu)$.  Let us call the result of this reconstruction
$A_{exp}(s|tu)$ and the corresponding $T,U,V$ amplitudes given by
Eqs.~(3.4) $T_{exp}, U_{exp}$ and $V_{exp}$ respectively.  We would
like to compare the experimental amplitude $A_{exp}$ with the
theoretical amplitude $A_{th}$ given in Sec. IV {\it in a whole
low-energy domain of the s-t-u plane including the unphysical region}.
Such a comparison should lead to a detailed fit, which in turn should
provide a rather precise determination of low energy constants
entering $A_{th}$.  In particular, we would like to measure the
parameter $\alpha$ and, in this way, let Nature tell us whether it
prefers a quark mass of effective dimension 1 or 2.  The theorem
proved in Sec. III considerably simplifies the above task:
Neglecting $O(p^8)$ contributions, the equation
\begin{equation}
  A_{exp} (s|tu) - A_{th} (s|tu) = 0,
\end{equation}
which is supposed to hold in a crossing symmetric domain of the
Mandelstam plane, is actually equivalent to a set of three {\it
single-variable} equations,
\begin{eqnarray}
T_{exp} (s) - T_{th} (s) &=& \delta T(s) \nonumber \\
U_{exp} (s) - U_{th} (s) &=& \delta U(s) \nonumber \\
V_{exp} (s) - V_{th} (s) &=& \delta V(s)
\end{eqnarray}
valid in an interval of $s$.  The functions $\delta T(s), \delta U(s)$
and $\delta V(s)$ are the arbitrary and irrelevant polynomials
given by Eq.~(3.5).  In this section, we will analyze Eqs.~(5.2).
Hereafter we systematically set $M_\pi^2 = 1$.

\pagebreak
\subsection{\bf One-loop precision}

The functions $T_{exp}, U_{exp}$ and $V_{exp}$ are given by Eqs.~(3.6).
Up to and including (one-loop) order $O(p^4)$, the theoretical
amplitude reads
\begin{equation}
T_{th} = T^{(0)} + T_{lead}^{(1)}
\end{equation}
(and likewise for $U$ and $V$), where the tree and leading one-loop
contributions are presented in Eqs.~(4.27) and (4.37) respectively.  We
shall concentrate on real parts of Eqs.~(5.2).

Let us denote the partial wave integrals appearing in Eqs.~(3.6) as
\begin{eqnarray}
    \phi_a(s) &=& \frac{s^3}{\pi} {\int\hspace{-1em}-}_4 ^{\Lambda^2}
    \frac{dx}{x^3} \frac{I\!m\,f_a(x)}{x-s}, ~~~~~~a = 0,2 \nonumber \\
    \phi_1(s) &=& \frac{s^2}{\pi} {\int\hspace{-1em}-}_4^{\Lambda^2}
    \frac{dx}{x^2} \frac{1}{x-4} \frac{I\!m\, f_1(x)}{x-s} ,
\end{eqnarray}
It is convenient to take linear combinations of the Eqs.~(5.2) for $T$
and $U$  and isolate the contributions of $I=0$ and $I=2$ $s$ waves.
The resulting equations can be written as $(a = 0,2)$
\begin{equation}
\phi_a(s) = \frac{\hat{\beta}^2N_a}{6\pi} (s + \kappa_a)^2
    D(s) + p_a (s),
\eqnum{5.5a}
\end{equation}
where \addtocounter{equation}{1}
\begin{equation}
 N_0 = 36,~~~~~~~~~~~~N_2 = 9
\end{equation}
and ($w \equiv \left\vert 1 - \frac{4}{s}\right\vert ^{1/2}$)
\begin{eqnarray}
  D(s) &\equiv& 6\pi~Re\,L(s), \nonumber \\
  D(s) &=& 12 + 6 w \ln \left\vert \frac{1-w}{1+w}\right\vert,~~~~s \leq 0, ~~s
\geq 4, \nonumber \\
  D(s) &=& 12 - 12 w \arctan w^{-1},~~~~ 0 \leq s \leq 4.
\end{eqnarray}
The $p_a(s)$ are two third-order polynomials, whose coefficients are
given in terms of (i) three constants $t_i$ [cf. the first of
Eqs.~(3.6)], (ii) the parameters $\alpha, \beta, \alpha_0(0)$ and
$\alpha_4(0)$ defined in terms of ${\cal L}_{eff}$, and (iii) the
irrelevant five constants that characterize the polynomial
ambiguity (3.5). The explicit expression for the coefficients of
$p_a(s)$ can be easily read off from Eqs.~(3.6), (4.27) and (4.37).
Similarly, the $V$-equation (5.2) can be written as
\begin{equation}
\phi_1(s) = \frac{\hat{\beta}^2}{6\pi} (s-4) D(s) + q(s),
\eqnum{5.5b}
\end{equation}
where $q(s)$ is now a second order polynomial with coefficients
given by linear combinations of three parameters $v_i$ [cf. the last
of Eqs.~(3.4)], the ${\cal L}_{eff}$  parameters $\beta_0(0)$ and
$\beta_2(0)$ and the irrelevant constants $y_i$.  For small $s$, the
function $D(s)$ behaves as
\begin{equation}
  D(s) = s + \frac{1}{10} s^2 + O(s^3) .
\end{equation}
On the other hand, the functions $\phi_a(s)$ and $\phi_1(s)$ defined
in (5.4) behave as $O(s^3)$ and $O(s^2)$ respectively. The polynomials
$p_a(s)$ and $q(s)$ should be such to insure this small s
behavior on the right hand sides of Eqs.~(5.5a) and (5.5b). Using
(5.8), one easily finds
\begin{eqnarray}
   p_a(s) &=& \frac{\hat{\beta}^2 N_a}{6\pi} \left\{ -\kappa_a^2 s -
\frac{1}{10} \kappa_a(\kappa_a + 20)s^2 + \tau_a s^3 \right\}~~~~a=0,2
\nonumber \\
   q(s) &=& \frac{\hat{\beta}^2}{6\pi}\left\{ -s(s-4) + \tau_1 s^2
\right\},
\end{eqnarray}
where $\tau_0,\tau_1,\tau_2$ are three yet undetermined parameters.
Eqs.~ (5.5a) and (5.5b) now take the form $(a=0,2)$
\begin{eqnarray}
\phi_a(s) &=& \frac{\hat{\beta}^2 N_a}{6\pi} \left\{ s^2 D(s) +
2s[D(s) -s] \kappa_a + [D(s) - s - \frac{1}{10} s^2]\kappa_a^2 +
\tau_a s^3 \right \} \nonumber \\
\phi_1(s) &=& \frac{\hat{\beta}^2}{6\pi} \left\{ (s-4) [D(s) - s]
+ \tau_1 s^2 \right\} .
\end{eqnarray}
Once the experimental phase shifts are known, one can compute the
integrals $\phi(s)$ on left-hand side of Eq.~(5.10) and fit them with
the corresponding right-hand side.  The parameters of the fit are
$\alpha, \beta, \tau_0, \tau_1, \tau_2$.  At this stage, one does not
need to know the subtraction constants $t_i$ and $v_i$ in the
dispersion relations (3.6).  The latter are needed, however, if one
wants to measure the four parameters of ${\cal L}^{(4)}$, namely
$\alpha_0(0), \beta_0(0), \delta\hat{\alpha}+\alpha_4(0)$ and
$\beta_2(0)$.  (Remember that the parameters $\alpha_0(0)$ and
$\beta_0(0)$ determine the two linear combinations (4.42) of the
low-energy constants $L_1, L_2$ and $L_3$ that appear in the ${\cal
L}_{40}$-part (4.31) of ${\cal L}_{eff}$.)  Indeed, comparing
coefficients of polynomials on both sides of Eqs.~(5.9), one gets 11
linear relations among the ``experimental" constants $t_0, t_2, t_3,
v_1, v_2, v_3$, the four parameters of ${\cal L}_{eff}$ mentioned
above, and the irrelevant five constants $x,y_0,y_1,y_2,y_3$.
Eliminating the latter, one can express the four ${\cal L}_{eff}$
parameters as
\begin{mathletters}
\begin{eqnarray}
   \alpha_0(0) &=& t_2 - \frac{\hat{\beta}^2}{10\pi} \left\{
     2\kappa_0(\kappa_0 + 20) + \kappa_2(\kappa_2 + 20) \right\}
\nonumber \\
   \beta_0(0) &=& v_2 + \frac{9\hat{\beta}^2}{\pi}
     (1 - 8\tau_0 + 5\tau_2) \nonumber \\
    &+& \frac{3\hat{\beta}^2}{40\pi} \left\{ 8\kappa_0 (\kappa_0 + 20)
      - 5\kappa_2(\kappa_2 + 20)\right\},
\end{eqnarray}
and
\begin{eqnarray}
\delta \hat{\alpha}+ \alpha_4(0) &=& t_0 - \hat{\alpha} -
\frac{4\hat{\beta}^2}{3\pi}
      (2\kappa_0^2 + \kappa_2^2) \nonumber \\
   \beta_2(0) &=& v_1 - 9\hat{\beta} + \frac{21\hat{\beta}^2}{20\pi}
    \left\{ 5\kappa_2^2 - 8 \kappa_0^2 \right\} \nonumber \\
    &+& \frac{6\hat{\beta}^2}{\pi} \left\{ 5(\kappa_2 - 2\tau_2)
     -8 (\kappa_0 - 2\tau_0) \right\} .
\end{eqnarray}
\end{mathletters}
The remaining two equations do not involve any parameter of ${\cal
L}_{eff}$ to be determined.  They read
\begin{eqnarray}
   t_3 &=& - \frac{\hat{\beta}^2}{\pi} (2\tau_0 + \tau_2) \nonumber \\
   v_3 &=& \frac{9\hat{\beta}^2}{4\pi} (1 - 8\tau_0 + 5\tau_2 - \tau_1).
\end{eqnarray}

The two Eqs.~(5.12) should be merely expected to measure the strength
of neglected two-loop and $\tilde{\cal L}^{(6)}$ contributions,
rather than represent a true constraint on the fit based on
Eqs.~(5.10).

\subsection{\bf Fits to Roy-type equations (3.7a) and (3.7b)}

In order to reconstruct the amplitude $A_{exp}(s|tu)$, one needs a
complete set of pion-pion phase shifts $\delta_a(s),~(a=0,1,2)$.  (By
complete we mean that they extend in energy from the threshold to
$\Lambda \lesssim 1~GeV$ for all three isospins and are dense enough in the
interval to allow adequate numerical evaluation of our dispersion
integrals.)

There exists only one complete set of pion-pion scattering phase
shifts (extrapolated from experimental data\footnote{For a recent
review of experimental $\pi - \pi$ scattering data, see
\cite{ochs-newsletter,gm}.})
that has been published in numerical form, namely that appearing in
the paper of Froggatt and Petersen \cite{fp}.  They provide values for
$\delta_a(s)$ --- without quoted errors --- at 20 MeV energy intervals
in $4M_\pi^2 < s < \Lambda^2$, for $a = 0,1,2$.  The phase shifts
$\delta_a$ come from an analysis following that of Basdevant {\it et
al.} \cite{bfp}, which employs a truncated set of twice-subtracted Roy
equations, makes a particular choice of parametrization for $f_a$
(fixing the $I=0$ scattering length, $a_0$) and uses a Regge type
model for estimating the high energy contributions to the dispersion
integrals. Data were taken from the Estabrooks-Martin analysis
\cite{em74} of the CERN-Munich experiment on $\pi N \rightarrow \pi
\pi N$ \cite{cern-munich}.  Although Basdevant {\it et al.} \cite{bfp}
present graphical results for several choices of values of $a_0$ in
their work, numerical results are only presented in the subsequent
paper of Froggatt and Petersen \cite{fp}, and only for the unique
choice $a_0 = 0.3$.

We first check to what extent the Froggatt-Petersen phases satisfy
the version of the Roy equations set forth in Section III B.  To this
end, we compute the integrals on the right hand side of Eqs.~(3.7a,b),
using the $\delta_a$ from Froggatt and Petersen.  Calling the result
$Ref_a^{RHS}(s)$, we then determine the parameters $t_i,v_i$ by minimizing
\begin{equation}
\sum_a \sum_i [ Re f_a^{LHS}(s_i) - Re f_a^{RHS} (s_i) ]^2,
\end{equation}
where $Ref_a^{LHS}$ is the real part of $f_a$ determined directly (via
unitarity) from $\delta_a$.  This is not a proper $\chi^2$ fit, since
no uncertainties can be included; consequently, no uncertainties can
be quoted for the resulting constants. We find, however, that the
values for experimentally determined constants are stable for
reasonable variations in the energy interval used for the fit (see
Table \ref{tuv}). The fit over the largest range, $4 < s < 25$, is
excellent: $Ref_a^{RHS}$ and $Ref_a^{LHS}$ agree to 1\% over nearly
all the interval, the sum in Eq.~(5.13) being $O(10^{-4})$ for 63 data
points.  We see no need to present the results graphically:
$Ref_a^{RHS}$ and $Ref_a^{LHS}$ would be indistinguishable.  Instead,
the values of $Ref_a^{RHS}$ and $Ref_a^{LHS}$ are compared in Table
\ref{fptable}, for 21 energies included in the sum (5.13).  We thus
conclude that the Froggatt-Petersen phases indeed give a solution of
our set of triply-subtracted Roy equations, for the values of
parameters $t_i$ and $v_i$ summarized in Table \ref{tuv}.  (Notice
that the parameters $t_3$ and $v_3$ are poorly determined, but that
they are sufficiently small not to affect the analysis at the $O(p^4)$
level.)  The corresponding low-energy amplitude $A_{exp}(s|tu)$ will
be confronted with the theoretical prediction $A_{th}$ shortly.

$K_{e4}$-decay experiments \cite{ke4} are consistent with the value
$a_0 = 0.30$ for the scattering length, characteristic of
Froggatt-Petersen phases, but standard $\chi$PT predicts a lower value,
namely $a_0 = 0.20 \pm 0.01$ \cite{gl84}.  It would be desirable to
have complete sets of phase shifts that fit both experiment and Roy
equations for other values of $a_0 < 0.30.$ These are not
available.\footnote{J.L.  Basdevant, private communication.} For this
reason, we must use {\it ad hoc} extrapolations down to threshold of
existing data at energies $E > (500-600)~ MeV$ obtained from $\pi N
\rightarrow \pi\pi N$ and $\pi N \rightarrow \pi\pi\Delta$ production
experiments.  One such extrapolation has been recently considered by
Schenk \cite{schenk} using a simple parametrization
\begin{eqnarray}
\tan \delta_i(s) &=& \sqrt{\frac{s-4}{s}} \left[ a_i + \tilde{b}_i
( \frac{s - 4}{4} ) + c_i ( \frac{s-4}{4})^2 \right]
(\frac{4-s_i}{s-s_i} ) \nonumber \\
    \tilde{b}_i &=& b_i - a_i \frac{4}{s_0 - 4} + (a_i)^3.
\end{eqnarray}
for the two $s$ waves ($i=0.2$) and a similar formula for the $p$
wave.  The scattering lengths $a_i$ and the slope parameters $b_i$ are
fixed at their values predicted by the standard one-loop $\chi$PT
\cite{gl85}:
\begin{eqnarray}
a_0 \equiv a_0^0 = 0.20, & a_2 \equiv a_0^2 = -0.042, & a_1 \equiv a_1^1 =
0.037 \nonumber \\
b_0 \equiv b_0^0 = 0.24, & b_2 \equiv b_0^2 = -0.075 & .
\end {eqnarray}
The remaining parameters are determined by fitting the data obtained
from various analyses of dipion production experiments
\cite{cern-munich}.  For the $I=0$ $s$ wave, Schenk uses the Ochs
energy-independent analysis\footnote{The data by Ochs can be found in
his unpublished thesis \cite{ochs}. We are indebted to Dr. J. Gasser
for communicating these unpublished data to us.  For the reader's
convenience, they are reproduced in our Table \ref{ochsdata}.} of the
CERN-Munich experiment \cite{em74}, covering the energy range 610--910
MeV.  For his best fit -- called solution B -- no $\chi^2$ or error
bars are quoted.  Instead, two additional sets of parameters $c_0$ and
$s_0 = E_0^2$ called ``A'' and ``C'' are given that bracket together
both the Ochs data and the well-known data by Estabrooks and Martin
\cite{em74}.  A similar procedure is adopted for the $I=2$ $s$ wave,
whereas the parameters of the $p$ wave are determined from the
experimental $\rho$ mass and width.  Results of this analysis and more
details can be found in Ref.\ \cite{schenk}.

In this way, the parametrization (5.14) provides a complete set of
phases -- hereafter referred to as Schenk B -- that fits the data at
higher energies and uses the threshold parameters (5.15) of the
standard $\chi$PT.  Using this set, we have performed exactly the same
kind of fit to the Roy-type Eqs.~(3.7a) and (3.7b) as in the case of
Froggatt-Petersen phases.  Surprisingly enough, we find this fit at
least as good as in the case of the Froggatt-Petersen phases, despite
the fact that the Schenk B phases were not obtained using Roy
equations or any other crossing-symmetry correlation among the three
lowest partial waves.\footnote{A. Schenk, private communication.} The
resulting parameters $t_i$ and $v_i$ are given in the second half of
Table \ref{tuv}, and the quality of the fit can be appreciated from
Table \ref{schenkb}.  Unfortunately, we do not see any simple way to
associate the Schenk B phases, and the corresponding parameters $t_i$
and $v_i$, with a set of errors which would be deduced from
statistical errors of the experimental data used at the beginning and
which would respect the correlations imposed by the Roy equations. The
same remark applies to the set of phases of Froggatt and Petersen.

\subsection{\bf Determination of parameters $\alpha,\beta,L_1,L_2$ and
$L_3$ \protect \\
from a complete set of phase shifts}

The next step is to confront the empirical amplitude $A_{exp}$ with
the amplitude $A_{th}$ computed from chiral perturbation theory.  In
particular, the two solutions of the Roy-type equations (3.7a) and
(3.7b) described above can be used to measure the parameters
$\alpha,\beta$ and, through Eqs.~(4.42), two linear combinations of the
low-energy constants $L_1^r,L_2^r$ and $L_3$, defining the
four-derivative terms in ${\cal L}_{eff}$.  The measurement is based on
Eqs.~(5.10).  First, one evaluates the three functions
$\phi_a(s),~a=0,1,2$, defined in Eqs.~(5.4), using the complete sets of
phase shifts exhibited in Section V B.  The results are represented
graphically by continuous lines in Figs. 3a,b,c for the case of
Froggatt-Petersen phases and in Figs. 4a,b,c for the Schenk B set.
Next, one fits the ``experimental'' functions $\phi_a(s)$ with the
theoretical expression represented on the right-hand side of
Eqs.~(5.10).  The parameters of the fit are $\alpha,\beta$ and
$\tau_0,\tau_1,\tau_2$. [Recall that the $\kappa_a$ are defined in
terms of the ratio $\alpha/\beta$ -- see Eq.~(4.34).]

The range in $s$ in which the fit is performed should not exceed the
range in which the $O(p^4)$-order $\chi$PT may actually be expected to
apply.  On the other hand, this range should be large enough to permit
a sensitive determination of parameters.  For this reason, it might be
misleading to consider exclusively the physical region $s \geq 4$
\cite{drv,gm}.
In the following, we use the interval $-4 \leq s \leq 8$, which most
likely represents a rather conservative choice.

{}From Figs. 3a,b,c and 4a,b,c one observes a large difference in
scale of individual $\phi_a$:  $\phi_0$ is typically an order of
magnitude or more larger than $\phi_2$ and nearly two orders of
magnitude larger than $\phi_1$.  For this reason, we first fit the
function $\phi_0$, determining the three parameters $\alpha,\beta$ and
$\tau_0$.  Then, using the values of $\alpha$ and $\beta$ obtained in
this way, we perform two single-parameter fits to $\phi_2$ and
$\phi_1$, determining $\tau_2$ and $\tau_1$ respectively.  In the
absence of error bars for $\phi_a(s)$, it is impossible to perform a
true $\chi^2$ fit.  Instead, we minimize the sum of squares of the
difference between the left- and right-hand sides of Eqs.~(5.10), for
66 equidistant points in the interval $-4 \leq s \leq 8$, giving the
same weight to each point.

In all cases, the parameter $\beta = F_\pi^2/F_0^2$ should remain
close to 1, and the fit should be constrained by this condition.  We
require
\begin{equation}
\beta \leq 1.17,
\end{equation}
corresponding to the lower bound $F_0 \geq 86~MeV$.  This bound is
consistent both with existing standard $\chi$PT estimates \cite{gl85} and
with the improved $\chi$PT formula (4.20).  Leaving the ratio
$\alpha/\beta$ unconstrained in the minimization procedure, one tests
-- for a given set of data -- the relevance of the improved $\chi$PT.  The
corresponding fits are represented by dashed curves in Figs. 3a,b,c
and 4a,b,c. The corresponding best values of the parameters are
\begin{mathletters}
\begin{eqnarray}
\alpha/\beta &=& 4.20, ~~~~~~ \beta = 1.17  \nonumber \\
\tau_0 &=& -0.263, ~~ \tau_1 = 3.75, ~~~ \tau_2 = -0.540
\end{eqnarray}
for the set of Froggatt-Petersen phases, and
\begin{eqnarray}
\alpha/\beta &=& 1.63, ~~~~~~ \beta = 1.17  \nonumber \\
\tau_0 &=& -0.032, ~~ \tau_1 = 3.68, ~~~ \tau_2 = -0.640
\end{eqnarray}
\end{mathletters}
for phases of the Schenk B set.  On the other hand, in order to test
the compatibility of the $O(p^4)$ {\it standard} $\chi$PT with a given set
of data, one further restricts the fit by requiring
\begin{equation}
\alpha = \beta \leq 1.17.
\end{equation}
Results of the minimization with this constraint are represented by
dot-dashed curves in Figs. 3 and 4.  The best values of parameters
corresponding to this constrained fit are
\begin{mathletters}
\begin{equation}
\alpha = \beta = 1.17, ~~~\tau_0 = -0.414,~~~\tau_1 = 3.75,~~~\tau_2
=-0.661
\end{equation}
and
\begin{equation}
\alpha = \beta = 1.17, ~~~\tau_0 = -0.045,~~~\tau_1 = 3.68,~~~\tau_2 =
-0.653
\end{equation}
\end{mathletters}
for the Froggatt-Petersen and Schenk B sets of phases respectively.

A few remarks are in order.  The Froggatt-Petersen data are
considerably better fit in terms of a larger value (5.17a) of the
ratio $\alpha/\beta$ than the standard $\chi$PT would permit, although
without a true $\chi^2$ fit we cannot be quantitative about this
observation.  The failure of standard $\chi$PT to describe the
Froggatt-Petersen $s$ wave is also apparent in Fig. 3a (dot-dashed
curve).  Concerning the $p$ wave, the fit is reasonably good for both
cases (Fig. 3b), reflecting the fact that the theoretical calculation
of $\phi_1(s)$ senses the effective infrared dimension of the quark
mass starting only at the two-loop level.  It is worth noting that the
best value $\alpha/\beta$ = 4.20 is overcritical by 5\%.  This means
that the Froggatt-Petersen I=0 $s$ wave would be compatible with the
vanishing of the $\bar{q}q$ condensate $B_0$.\footnote{This critical
case has been considered earlier \cite{mass-dimension}.} From this
point of view, the set of Froggatt-Petersen phases with $a_0^0 = 0.30$
appears as an extreme alternative.  The opposite extreme is
represented by the Schenk B set of phases.  Since the latter
incorporates {\it a priori} the values of scattering lengths and
effective ranges as predicted by the standard $\chi$PT, it is not
surprising that the corresponding best value for $\alpha/\beta$
(5.17b) is considerably closer to 1 than in the Froggatt-Petersen
case.  Furthermore, Fig. 4a seems to indicate that, although the best
value for $\alpha/\beta$ is still as large as 1.63, this fact need not
be significant.  In the absence of error analysis, it is hard to be
too affirmative in the interpretation of the Schenk B fit.

It remains to exploit the additional information (values of constants
$t$ and $v$ as well as the constants $\tau$ resulting from our fits),
in order to measure certain parameters of the dimension-4 component of
${\cal L}_{eff}$.  Here, we merely concentrate on the constants
$L_1,L_2$ and $L_3$ characteristic of ${\cal L}_{40}$, Eq.~(4.31),
whose meaning and renormalization do not depend on the effective
dimension of the quark mass.  For this purpose, we have to determine
the constants $\alpha_0(0)$ and $\beta_0(0)$ given by Eqs.~(5.11a).
Using the central values of the parameters $t_2$ and $v_2$ (the second
column of Table \ref{tuv}) and the best values for $\alpha/\beta$ and
the $\tau$'s, as determined in the previous fits, one gets
$\alpha_0(0) = 5.81 \times 10^{-4}, \beta_0(0) = 3.99 \times 10^{-3}$
for the Froggatt-Petersen solution, and $\alpha_0(0) = 5.87 \times
10^{-4}, \beta_0(0) = 2.07 \times 10^{-3}$ for the case of Schenk B
phases.  These numbers are easily converted into information on the
constants $L_{1,2,3}$, using Eqs.~(4.42) and (4.43).  Assuming the
Zweig-rule (or large-$N_c$) relation $L_2^r - 2L_1^r = 0$, and
identifying the running scale $\bar{\mu}$ with the $\eta$ mass, as
done in Refs.\ \cite{gl85,rigg}, one obtains
\begin{mathletters}
\begin{equation}
L_2^r = 2L_1^r = 1.34 \times 10^{-3},~~~L_3 = -4.50 \times 10^{-3}
\end{equation}
for the Froggatt-Petersen data, and
\begin{equation}
L_2^r = 2L_1^r = 0.56 \times 10^{-3},~~~L_3 = -2.15 \times 10^{-3}
\end{equation}
\end{mathletters}
for the set of Schenk B phases.  It is gratifying to see that these
values -- especially (5.20a) -- compare well with other determinations
based on {\it standard} $\chi$PT \cite{gl85,rigg}.  Indeed, there is no
reason why the purely derivative terms in ${\cal L}_{eff}$ should be
affected by questions concerning the symmetry breaking sector.

\subsection{\bf Estimates of errors in the direct measurements of
$\alpha/\beta$}

The uncertainties in the values of the parameters $\alpha,\beta$ and
$\tau_a$ arise from uncertainties in the functions $\phi_a$; these
uncertainties, in turn, arise from uncertainties in the phase shifts
$\delta_a$ over the range of integration in Eqs.~(5.4).  As we have
noted, there is no set of phase shifts $\delta_a$ which exists,
together with corresponding errors, in this energy range.  In the
present subsection, we extend the extrapolation method of Schenk
\cite{schenk}, described above, to construct several sets of I=0 phase
shifts $\delta_0$, together with estimated errors, in the necessary
energy interval.  In this way, we obtain values and estimated errors
for the parameters $\alpha/\beta$ and $\tau_0$ for each extrapolated
data set.  Only $I=0$ phases are considered. In fact, we could treat
$I=1$ phases similarly (although the insensitivity of $\phi_1$ to
$\alpha$ makes this relatively uninteresting); in any case, the
paucity of experimental data on $I=2$ makes the production of a
complete set of phase shifts impossible without a more extensive
recourse to the use of the Roy equations, as in the analysis of
Basdevant {\it et al.} \cite{bfp}.

The two original sets of phase shifts (with corresponding errors) used
are that of Ochs and that of Estabrooks and Martin.  These were each
obtained independently from analysis of the same CERN-Munich
experiment.  The first step is to extrapolate $\delta_0$ down to
threshold, using the Schenk formula (5.14).  The Ochs phases are fit
over the energy range 610-910 MeV, {\it i.e.}, using all his data for
which no inelasticity is suggested.  (See Table \ref{ochsdata}.)  The
Estabrooks-Martin phases are fit over the energy range 570-910 MeV,
{\it i.e.}, using all their points in the elastic scattering region
except for their first three lowest-energy points, which appear to be
less trustworthy.  In performing the extrapolation, the scattering
length $a_0$ is fixed and the remaining parameters $b_0,c_0,E_0$ are
determined by minimization of $\chi^2$ using the phases and errors
given to us.  We show in Fig. \ref{fits} the results of this fitting
procedure for the choices $a_0$ = 0.20 (preferred by standard $\chi$PT)
and 0.26 (preferred by $K_{e4}$-decay experiment) for each of the two
data sets; the resulting parameters are given in columns 2--5 of Table
\ref{phi0fits}.  The $\chi^2$ for these fits is quite good.  (We note in
passing that the data of Estabrooks and Martin is not well described
by the parameters $a_0=0.20,b_0=0.24$ which characterize the Schenk B
solution.)  The next step is to estimate the uncertainty in the
extrapolated phases $\delta_0$.  Since the dominant parameter (after
$a_0$, which is fixed) is $b_0$, and in view of the strong
correlations among the parameters, we proceed as follows: for fixed
values of $b_0$ larger than its value for $\chi^2_{min}$, the
minimum-$\chi^2$ value, fit the data by allowing $c_0$ and $E_0$ to
vary freely, and find the values of $b_0,c_0,E_0$ which give $\chi^2 =
\chi^2_{min} +1$; call this solution ``a'', in analogy with Schenk's
notation; repeat this procedure for fixed values of $b_0$ smaller than
that for $\chi^2_{min}$; call this solution ``c''; the uncertainty in
the phase shift $\delta_0(E_i)$, for each value of $E=E_i$, is then
estimated by interpreting the variation of $\delta_0(E_i)$ from its
solution ``a'' value to its solution ``c'' value as $\pm 1$ standard
deviation in $\delta_0 (E_i)$.  (This is similar to the procedure
adopted by Schenk, although he allows much greater variation --
leading to much larger uncertainties -- in order to bracket both Ochs
and Estabrooks-Martin phases at the same time.)

Now, for each of the four sets of phase shifts $\delta_0$, obtained by
the extrapolation procedure described above, we may make the
comparison for $\phi_0$ as done for the Froggatt-Petersen and Schenk B
phases in Section IV C.  However, we now have the important advantage
that a true $\chi^2$ fit is possible, so we can have some idea of the
precision with which the resulting parameters are determined.  For
each set, we make two fits: one, corresponding to standard $\chi$PT, for
which we fix $\alpha = \beta \leq 1.17$; the other, corresponding to
improved $\chi$PT, for which $\beta \leq 1.17$ but $\alpha$ is allowed to
vary freely.  The fits are all performed over the same interval $-7
\leq s \leq 9$.  Results of the determination of the parameters
$\alpha/\beta$ and $\tau_0$ are given in columns 6--9 of Table
\ref{phi0fits}; the reader can judge the quality of the fits from the
plots of $\phi_0$ given in Figs.\ref{phi0figs}a,b,c,d. The solid
curves represent the parametrization of improved $\chi$PT, while the
dashed curves represent that of standard $\chi$PT. It is clear, both from
the large $\chi^2$ values tabulated for the standard $\chi$PT fits and
from examination of the dashed curves in Figs.\ref{phi0figs} that
standard $\chi$PT is not compatible with these phase shifts.  For this
reason, we quote no result for $\tau_0$ for this case.  On the other
hand, improved $\chi$PT can easily accommodate such data.  It is important
to note that, for a given set of phase shifts, the parameters
$\alpha/\beta$ and $\tau_0$ are very well determined by the improved
$\chi$PT fit.

As a check on our procedure of estimating errors, we have also used a
more ``conservative'' procedure, {\it viz.}, vary all non-fixed
parameters within their one-standard-deviation limits to produce
solution ``$a_{cons}$'', taking max($b_0$),max($c_0$),min($s_0$), and
solution ``$c_{cons}$'', taking min($b_0$),min($c_0$),max($s_0$); then
compute the ``conservative'' uncertainties in the phase shifts
$\delta_0(E_i)$ using these solutions as we did for solutions ``a''
and ``c'' before.  Clearly, this method does not take into account the
strong correlations in $b_0,c_0,s_0$.  Thus, when the consequent phase
shift errors are used in the fitting of $\phi_0$, these larger errors
result in larger errors in the experimental function $\phi_0$.  The
result is then a $\chi^2$ roughly half of that previously obtained,
and errors in $\alpha/\beta$ and $\tau_0$ roughly 2--5 times larger.
Nevertheless, the best fit is the same.
\section{\bf Summary and Conclusions}

A new framework for testing the convergence rate of chiral
perturbation theory is proposed.  One first replaces the standard
expansion of the effective Lagrangian by a more general expansion
that is as systematic and unambiguous as the standard $\chi$PT.  In
addition to the usual terms, the new expansion involves at each given
order new contributions that the standard $\chi$PT relegates to
higher orders.  The size of these additional contributions can then be
tested experimentally, in particular in low-energy $\pi-\pi$
scattering.  Unless these contributions turn out to be small, the
improved $\chi$PT has, in principle, more chance to produce a rapidly
convergent expansion scheme.

A new low-energy theorem is presented which provides the general
solution of constraints imposed by analyticity, crossing symmetry and
unitarity on the $\pi-\pi$ scattering amplitude, neglecting $O(p^8)$
contributions.  Applications of this theorem are threefold:

i)  First, it considerably simplifies the evaluation of the
perturbative $\pi-\pi$ amplitude up to and including two loops.  This
applies both within the ``standard $\chi$PT'' and within the more
general ``improved $\chi$PT'' which contains the former as a special
case.  In both cases, the calculation reduces to the iterative
insertion of the unitarity condition (4.32) into the dispersive
integral for the functions $T,U$ and $V$ in Eq.(3.2).  The improved
$\chi$PT one-loop amplitude is worked out in detail in Sec. IV.  The
two-loop amplitude can be easily calculated along the same lines.  The
reason why the formula (3.2) no longer holds beyond two loops
resides in new $O(p^8)$ effects in the absorptive part:
inelasticities and higher partial waves.

ii)  Next, the low-energy theorem of Sec. III can be used to constrain
the low-energy scattering data and to fully reconstruct the
corresponding amplitude.  The formula (3.2) implies a particular
truncation of the infinite system of Roy equations, under a rigorous
control of chiral power counting:  Neglected contributions are
$O(p^8)$, whereas in the original form of the Roy equations \cite{roy}
the model-dependent ``driving terms'' are of the same order $O(p^4)$
as the effects we are looking for.  A complete set of low-energy
phases $\delta_0^0,\delta_0^2$ and $\delta_1^1$, together with the six
subtraction constants $t$ and $v$ for which the Roy-type Eqs. (2.7a)
and (2.7b) are satisfied to a reasonable accuracy (see Tables
\ref{fptable} and \ref{schenkb}),define up to $O(p^8)$corrections the
scattering amplitude $A(s,t,u)$ in a whole low-energy region of the
Mandelstam plane including the unphysical region.  Two examples of
such a complete low-energy amplitude are given, based on
phase shifts published by Froggatt and Petersen \cite{fp} and by
Schenk \cite{schenk} respectively.  They are both compatible with
existing $\pi N \rightarrow \pi\pi N$ and $K_{e4}$ experimental data.

iii)  Finally, the low-energy representation (3.2) simplifies the
direct comparison of the perturbative amplitude $A^{th}(s,t,u)$ with
the amplitude $A^{exp}(s,t,u)$ reconstructed from the data.  In
particular, parameters of ${\cal L}_{eff}$ contained in $A^{th}$ can
be measured through a detailed fit of the amplitude $A^{exp}(s,t,u)$
over a sufficiently large portion of the Mandelstam plane in which the
low-energy expansion can still be taken as valid.  The fit is
particularly sensitive to the ratio $\alpha/\beta$ which parametrizes
the leading $O(p^2)$ amplitude.  The improved $\chi$PT requires $1
\leq \alpha/\beta \leq 4$, whereas the special case of the standard
$\chi$PT corresponds to $\alpha/\beta = 1$.  The ratio $\alpha/\beta$
is related to the value of the QCD parameter $2\hat{m}B_0$ in the
units of pion mass squared and, {\it via} the pseudoscalar mass
spectrum, to the quark mass ratio $r = m_s/\hat{m}$.

Examples of measurement of $\alpha/\beta$ exhibited in this paper
illustrate the lack of sufficiently precise experimental information
on low-energy $\pi-\pi$ scattering.  For a fixed value of the
scattering length $a_0^0$, the statistical errors of the production
data on $\delta_0^0$ are estimated to show up as errors in the
measured values of $\alpha/\beta$ of the order of a few percent.  On
the other hand, for different values of $a_0^0$ in the experimental
range $a_0^0 = 0.26 \pm 0.05$, and for different sets of production
data, the resulting values of $\alpha/\beta$ vary between 1.5 and 4.2.
The two complete low-energy amplitudes mentioned above correspond to
these two extremes.  In particular, the Froggatt-Petersen phases (for
which $a_0^0$ = 0.30) are compatible with the vanishing of the
condensate $B_0$ and with the critical value of the quark mass ratio
$r = r_1 \simeq 6.3$.

The suspicion that a bad convergence of the standard $\chi$PT might
bias the usual conclusions that $r = m_s/\hat{m} \simeq 25.9$ and
$2\hat{m}B_0 \simeq M_\pi^2$ is at least well motivated, but clearly
it requires confirmation.  In order to produce a truly unbiased
measurement of these fundamental QCD parameters, the method developed
in this paper can prove useful provided that it is supplied with more
accurate experimental information on low-energy $\pi-\pi$ phase
shifts.  The current imprecision, illustrated by error bars as large
as in $a_0^0 = 0.26 \pm 0.05$ can hide {\it all} cases of interest,
including the intriguing critical case $\langle \bar{q}q \rangle = 0$.
Here, one faces a challenge of fundamental high precision low-energy
experimental physics.
\acknowledgments

We are indebted to Dr. M. Knecht for various discussions and for
pointing out some errors in a preliminary version.  Discussions on
$\pi-\pi$ data with Dr. J.-L. Basdevant and with Dr. J. Gasser at
various stages of this work have been extremely useful.  One of us
(N.H.F.) would like to thank the Division de Physique Th\'{e}orique,
Orsay, for its generous hospitality and support during his sabbatical
leave of absence, and the Department of Physics of Arizona State
University for its hospitality as well.
\appendix
\section{Notation and conventions}

In the first part of this appendix, we fix the notation and
normalization for the scattering amplitude.  We then exhibit the main
properties of the crossing matrices $C$ [Eq.~(3.9)].

The $S$-matrix element for the transition $\pi^a + \pi^b \rightarrow
\pi^c + \pi^d$, where $a,b,c,d$ are pion isospin indices, is connected
to the $T$-matrix element by the relation:
\begin{equation}
\langle cd | S | ab \rangle = \langle cd | ab \rangle +
i(2\pi)^4\,\delta^4(p_a +p_b - p_c -p_d)\,T_{ab,cd}
\end{equation}
The $T$-matrix element can be written in terms of isospin invariant
amplitudes; taking  crossing symmetry into account, the decomposition
reads:
\begin{eqnarray}
T_{ab,cd}(s,t,u) &=& A(s|tu) \delta_{ab}\delta_{cd} + A(t|su)
\delta_{ac} \delta_{bd} + A(u|ts) \delta_{ad}\delta_{bc},
\end{eqnarray}
where $s,t$ and $u$ are the Mandelstam variables,
\begin{equation}
s = (p_a + p_b)^2,~~~t = (p_a - p_c)^2,~~~u = (p_a - p_d)^2.
\end{equation}
The amplitude $A(s|tu)$ is symmetric in the variables $t,u.$  (The
amplitude $A(t|su)$ is obtained from $A(s|tu)$ by the exchange of
variables $s,t$ and by subsequent analytic continuation.)

The $s$-channel isospin amplitudes $F^{(I)}$ [Eq.~(3.8)] are related to
the amplitude $A$ by
\begin{equation}
\left(
\begin{array}{c}
F^{(0)}\\F^{(1)}\\F^{(2)}
\end{array}
\right) (s,t,u) = \frac{1}{32\pi}\left(
\begin{array}{rrr}
3 &~~ 1 &~~ 1 \\
0 &~~ 1 &~~ -1 \\
0 &~~ 1 &~~ 1
\end{array}
\right)
\left(
\begin{array}{c}
A(s|tu)\\A(t|su)\\A(u|ts)
\end{array}
\right).
\end{equation}
The partial wave expansion is
\begin{equation}
F^{(I)} = \sum_\ell (2\ell +1) P_\ell (\cos \theta) f_\ell^I(s),
\end{equation}
where $s = 4(M_\pi^2 + q^2), t = -2q^2(1-\cos\theta)$.  With this
normalization, the elastic unitarity condition for $f_\ell^I$ takes
the form:
\begin{mathletters}
\begin{eqnarray}
I\!m\, f_\ell^I(s) &=& \sqrt{\frac{s-4M_\pi^2}{s}} |f_\ell^I(s)|^2,\\
f_\ell^I(s) &=& \sqrt{\frac{s}{s-4M_\pi^2}} e^{i\delta_\ell^I(s)} \sin
\delta_\ell^I(s).
\end{eqnarray}
\end{mathletters}

The crossing matrices $C$ [Eq.~(3.9)] have the following forms:
\widetext
\begin{equation}
C_{st} = \left(
\begin{array}{rrr}
1/3 &~~1&~~5/3 \\
1/3 &~~1/2&~~-5/6 \\
1/3 &~~-1/2&~~1/6
\end{array}
\right),
C_{tu} = \left(
\begin{array}{rrr}
1 &~~0&~~0 \\
0 &~~-1&~~0 \\
0 &~~0 &~~1
\end{array}
\right),
C_{su} = \left(
\begin{array}{rrr}
1/3 &~~-1&~~5/3 \\
-1/3 &~~1/2&~~5/6 \\
1/3 &~~1/2&~~1/6
\end{array}
\right),
\end{equation}
\narrowtext
and satisfy the relations
\begin{equation}
C_{tu}^2 =C_{su}^2 =C_{st}^2 =1,
\end{equation}
\begin{eqnarray}
C_{st}C_{su} & = C_{tu}C_{st} & = C_{su}C_{tu},\nonumber \\
C_{su}C_{st} & = C_{tu}C_{su} & = C_{st}C_{tu} .
\end{eqnarray}
It is worthwhile to notice that Eqs. (A9) imply that the eigenvectors
$A_{\pm}$ with eigenvalues $\pm 1$, respectively, of the matrix
$C_{tu}$ satisfy
\begin{equation}
C_{su}C_{st}A_{\pm} = \pm C_{st}A_{\pm}.
\end{equation}
These relations are extensively used throughout the calculation of
Sec. III C.

The invariance of the amplitude $A(s|tu)$ under the exchange of the
variables $t,u$ permits us to construct rather easily all independent
crossing symmetric polynomials of a given degree in $s,t,u$.  It is
convenient to take the variables $t,u$ as independent.  Since there
are
\begin{equation}
k_n \equiv [n/2] + 1
\end{equation}
symmetric monomials in $t,u$ that are homogeneous of degree $n$, the
number of independent parameters in a general crossing symmetric
polynomial of degree $N$ is
\begin{equation}
K_N \equiv \sum_{n=0}^N k_n.
\end{equation}
Hence, the most general polynomial of degree three contains six
parameters, as claimed in Sec. III C.

\section{Ambiguities of the amplitudes $T,U$, and $V$}

In this appendix, we determine the general expression for the
transformations $T \rightarrow T + \delta T,~U \rightarrow U + \delta
U,$ and $V \rightarrow V + \delta V$ that leave invariant the
scattering amplitude $A$ [Eq.~(3.2)].  It follows that the variations
$\delta T,~\delta U$ and $\delta V$ must satisfy the equation
\widetext
\begin{equation}
\delta T(s) + \delta T(t) + \delta T(u) + \frac{1}{3}[2\delta U(s) -
\delta U(t) - \delta U(u)] + \frac{1}{3}[(s-t)\delta V(u) +
(s-u)\delta V(t)] = 0.
\end{equation}
\narrowtext
In order to solve Eq.~(B1), one notices that only two out of the three
variables $s,t,u$ are independent.  By successive differentiation with
respect to independent variables, one obtains a set of simpler
equations which can be solved easily.  To simplify notation, let us
define
\begin{equation}
f \equiv \delta T,~~g \equiv \delta U,~~h \equiv \delta V.
\end{equation}
We first consider $s$ and $t$ as independent variables, and
differentiate Eq.~(B1) first with respect to $s$ and then with respect
to $t$.  We thus obtain the following two equations, where the primes
indicate differentiation with respect to the arguments of the
functions:
\widetext
\begin{equation}
f'(s) - f'(u) + \frac{1}{3}(2g'(s) + g'(u)) + \frac{1}{3}(h(u) +
2h(t)) - \frac{1}{3}(s-t)h'(u) = 0,
\end{equation}
\begin{equation}
f''(u) - \frac{1}{3}g''(u) + \frac{2}{3}h'(t) + \frac{1}{3}(s-t)h''(u)
= 0.
\end{equation}
\narrowtext
We now consider $t$ and $u$ as independent variables and differentiate
Eq.~(B4) with respect to $t$, obtaining the result
\begin{equation}
h''(t) - h''(u) = 0,
\end{equation}
which indicates that $h''$ is a constant, and therefore $h$ is a
quadratic polynomial,
\begin{equation}
h(t) = \frac{1}{2}at^2 + bt + c,
\end{equation}
where $a,b,c$ are constants.  Using this result for $h$ in Eq.~(B4),
we find a relation between $f$ and $g$:
\begin{equation}
f(u) = \frac{1}{3}g(u) + \frac{1}{18}au^3 -\frac{1}{3}(b+2M_\pi^2a)u^2
+du +e,
\end{equation}
where $d$ and $e$ are constants.  We then return to Eq.~(B3), consider
$s$ and $u$ as independent variables, and differentiate with respect
to $s$.  This implies
\begin{equation}
f''(s) + \frac{2}{3}g''(s) -\frac{4}{3}b -\frac{2}{3}a(4M_\pi^2 -s) =
0,
\end{equation}
which becomes, after replacing $f$ in terms of $g$ [Eq.~(B7)],
\begin{equation}
g''(s) + as -2(b+2M_\pi^2 a) = 0,
\end{equation}
the solution of which is
\begin{equation}
g(s) = -\frac{a}{6}s^3 +(b+2M_\pi^2 a)s^2 +ks +\ell,
\end{equation}
where $k$ and $\ell$ are constants.  The expression for $f$ then
becomes
\begin{equation}
f(s) = (d + \frac{k}{3})s + (e+\frac{\ell}{3}).
\end{equation}
Finally, upon substituting the expressions for $h$ [Eq.~(B6)], $g$
[Eq.~(B10)] and $f$ [Eq.~(B11)] in the original equation (B1), we
obtain two constraints on the constants
\begin{eqnarray}
(e + \frac{\ell}{3}) &=& -\frac{4M_\pi^2}{3} (d + \frac{k}{3}),\\
c &=& -(k + 4M_\pi^2 b + \frac{16}{3}M_\pi^4 a).
\end{eqnarray}
After relabeling the constants as follows,
\begin{equation}
y_0 = \ell,~ y_1 = k,~ y_2 = (b+2M_\pi^2 a),~ y_3 = -a/6,~ x = d + k/3,
\end{equation}
the functions $f,g$ and $h$, and hence $\delta T,~\delta U$ and
$\delta V$ [Eqs. (B2)], take the forms given in Eqs. (3.5).

\section{Polynomials and kernels of the Roy type equations}

In this Appendix we explicitly list the polynomials $P_a(s)$ and the
kernels $W_{ab}$ which appear in the Roy-type dispersion relations
(3.7).
\begin{eqnarray}
P_0(s) &=& 5t_0 + \frac{5}{9}t_2 [3s^2 + 2(s-4M_\pi^2)^2]
       +\frac{5}{6}t_3 [2s^3 - (s-4M_\pi^2)^3] \nonumber \\
       & +& \frac{2}{9}v_1 (3s-4M_\pi^2)
-\frac{8}{27}v_2(s-M_\pi^2)(s-4M_\pi^2) \nonumber \\
       & +& \frac{1}{27}v_3(5s-4M_\pi^2)(s-4M_\pi^2)^2 \\
P_2(s) &=& 2t_0 + \frac{2}{9}t_2 [3s^2 + 2(s-4M_\pi^2)^2] + \frac{1}{3}t_3
[2s^3 - (s-4M_\pi^2)^3] \nonumber \\
       & -& \frac{1}{9}v_1 (3s-4M_\pi^2)
+\frac{4}{27}v_2(s-M_\pi^2)(s-4M_\pi^2) \nonumber \\
       & -& \frac{1}{54}v_3(5s-4M_\pi^2)(s-4M_\pi^2)^2 \\
P_1(s) &=&\frac{1}{9}(s-4M_\pi^2)(v_1 + v_2 s)  \nonumber \\
&+& \frac{2}{27}(s-4M_\pi^2)v_3 [s^2 - \frac{1}{20}(s-4M_\pi^2)(11s -
4M_\pi^2)]
\end{eqnarray}

{\widetext
\begin{eqnarray}
\frac{1}{\pi}\int_{4M_\pi^2}^{\Lambda^2} \frac{dx}{x} \sum_{b=0}^2
W_{0b}(s,x)  I\!m\, f_b(x) &=&\frac{4}{\pi}(s-4M_\pi^2)^2
(s-2M_\pi^2) \int_{4M_\pi^2}^{\Lambda^2}  \frac{dx}{x^3} \frac{I\!m\,
f_1(x)}{(x-4M_\pi^2)} \nonumber \\
 & - & \frac{1}{6\pi}(s-4M_\pi^2)^3 \int_{4M_\pi^2}^{\Lambda^2}
\frac{dx}{x^4} \{ I\!m\, f_0(x) + 5 I\!m\, f_2(x) \nonumber \\
 & + & 9\left( 1 + \frac{2s}{x-4M_\pi^2}\right) I\!m\, f_1(x) \}
G\left(\frac{s-4M_\pi^2}{x}\right)
\end{eqnarray}

\begin{eqnarray}
\frac{1}{\pi}\int_{4M_\pi^2}^{\Lambda^2} \frac{dx}{x} \sum_{b=0}^2
W_{2b}(s,x)  I\!m\, f_b(x) &=& -\frac{2}{\pi}(s-4M_\pi^2)^2
(s-2M_\pi^2) \int_{4M_\pi^2}^{\Lambda^2}  \frac{dx}{x^3} \frac{I\!m\,
f_1(x)}{(x-4M_\pi^2)} \nonumber \\
& -& \frac{1}{12\pi}(s-4M_\pi^2)^3 \int_{4M_\pi^2}^{\Lambda^2}
\frac{dx}{x^4} \{ 2I\!m\, f_0(x) +  I\!m\, f_2(x) \nonumber \\
& -& 9\left( 1 + \frac{2s}{x-4M_\pi^2}\right) I\!m\, f_1(x) \}
G\left(\frac{s-4M_\pi^2}{x}\right)
\end{eqnarray}

\begin{eqnarray}
\frac{1}{\pi}\int_{4M_\pi^2}^{\Lambda^2} \frac{dx}{x} \sum_{b=0}^2
W_{1b}(s,x)  I\!m\, f_b(x) &=& -\frac{1}{\pi}(s-4M_\pi^2)^2
(s-2M_\pi^2) \int_{4M_\pi^2}^{\Lambda^2}  \frac{dx}{x^3}\frac{I\!m\,
f_1(x)}{(x-4M_\pi^2)} \nonumber \\
&\hspace{-2cm} + &\hspace{-1cm}\frac{1}{12\pi}(s-4M_\pi^2)^2
\int_{4M_\pi^2}^{\Lambda^2}
\frac{dx}{x^3} [2 - \left(2 +
\frac{s-4M_\pi^2}{x}\right)G\left(\frac{s-4M_\pi^2}{x}\right) ]
\nonumber \\
&&\hspace{-2cm}\times \{ 2I\!m\, f_0(x) - 5 I\!m\, f_2(x) + 9\left( 1 +
\frac{2s}{x-4M_\pi^2}\right) I\!m\, f_1(x) \}
\end{eqnarray}
}In Eqs. (C4)-(C6), the function $G$ is defined as follows:
\begin{eqnarray}
G(x) & \equiv & 4 \int_0^1 dy \frac{y^3}{1+xy} = \frac{4}{3x} -
\frac{2}{x^2} + \frac{4}{x^3} -\frac{4}{x^4}\ln (1+x), \nonumber \\
G(0) & = & 1.
\end{eqnarray}

\begin{figure}
\caption{The region of allowed values for the ratios $\alpha/\beta$ and
$m_s/\hat{m}$ lies between the two curves shown.}
\end{figure}
\begin{figure}
\caption{Phase shift $\delta_0 - \delta_1$ from data sets of Froggatt
and Petersen \protect \cite{fp} (dashed curve) and Schenk B\ \protect
\cite{schenk} (solid curve) compared with experimental data\
\protect\cite{ke4} from $K_{e4}$-decay.}
\end{figure}
\begin{figure}
\caption{The functions $\phi_a$ (shown as solid curves) for (a) $a=0$,
(b) $a=1$ and (c) $a=2$, using experimental phase shifts given by
Froggatt and Petersen\ \protect \cite{fp}.  Comparison is made with theoretical
fits: those of the standard $\chi$PT are shown as dot-dashed curves, while
the improved $\chi$PT fits are shown as dashed curves.}
\end{figure}
\begin{figure}
\caption{The functions $\phi_a$ (shown as solid curves), using phase
shifts from the Schenk B \protect \cite{schenk} parametrization of the
phase shifts of Ochs \protect \cite{ochs}.  The meaning of the curves
is the same as in Figure 3.}
\end{figure}
\begin{figure}
\caption{(a) Schenk-type parametrization of phase shift
data of Ochs, fixing $a_0 = 0.20$ (solid curve) and $a_0 = 0.26$
(dashed curve); (b) Schenk-type parametrization of phase shift data of
Estabrooks and Martin, fixing $a_0 = 0.20$ (solid curve) and $a_0 =
0.26$ (dashed curve).  Details are given in the text.}\label{fits}
\end{figure}
\begin{figure}
\caption{The function $\phi_0$ (shown as points with error bars), using
Schenk-type parametrization of phase shift data: (a) data of Ochs,
fixing $a_0 = 0.20$; (b) data of Ochs, fixing $a_0 = 0.26$; (c) data
of Estabrooks and Martin, fixing $a_0 = 0.20$; (d) data of Estabrooks
and Martin, fixing $a_0 = 0.26$.  In each case, the solid curve
represents the parametrization of improved $\chi$PT, while the dashed
curve represents that of standard $\chi$PT.}\label{phi0figs}
\end{figure}

\mediumtext
\begin{table}
\caption{Parameters resulting from fitting Eqs.(3.7a,b)}\label{tuv}
\begin{tabular}{clll}
\multicolumn{1}{c}{Parameter} &
\multicolumn{3}{c}{Energy Range for Fit (MeV)}\\
\multicolumn{1}{c}{$~$} &
\multicolumn{1}{c}{300-580} &
\multicolumn{1}{l}{~300-640} &
\multicolumn{1}{l}{~300-700} \\
\tableline
\multicolumn{4}{c}{Using phase shifts of Froggatt and Petersen:}\\
\tableline
$t_0$ &\dec 0.0206 & 0.0207 & 0.0208 \\
$t_2$ &\dec  6.4 $\times 10^{-4}$ & 6.5 $\times 10^{-4}$ & 6.7 $\times
10^{-4}$\\
$t_3$ &\dec  5.2 $\times 10^{-6}$ & 3.5 $\times 10^{-6}$ & 1.6 $\times
10^{-6}$\\
$v_1$ &\dec  0.0764 & 0.0760 & 0.0755\\
$v_2$ &\dec  0.0021 & 0.0020 & 0.0020\\
$v_3$ &\dec  -1.3$\times 10^{-5}$ & -4.4$\times 10^{-6}$  &+3.5$\times
10^{-6}$\\
\tableline
\multicolumn{4}{c}{Using phase shifts of Schenk, solution B:}\\
\tableline
$t_0$ &\dec 0.0067 & 0.0065 & 0.0063 \\
$t_2$ &\dec  4.6 $\times 10^{-4}$ & 4.8 $\times 10^{-4}$ & 5.1 $\times
10^{-4}$\\
$t_3$ &\dec  1.4 $\times 10^{-5}$ & 9.9 $\times 10^{-6}$ & 6.9 $\times
10^{-6}$\\
$v_1$ &\dec  0.0697 & 0.0695 & 0.0693\\
$v_2$ &\dec  0.0021 & 0.0020 & 0.0020\\
$v_3$ &\dec  -8.4$\times 10^{-7}$ & 2.3$\times 10^{-6}$ & 6.5$\times 10^{-6}$\\
\end{tabular}
\end{table}

\mediumtext
\begin{table}
\caption{Comparison of the left and right hand sides of  Eqs.(3.7a,b),
using phase shifts of Froggatt and Petersen}\label{fptable}
\begin{tabular}{c|cc|cc|cc}\hline
\multicolumn{1}{c}{$~$} &
\multicolumn{2}{c|}{I=0} &
\multicolumn{2}{c|}{I=1} &
\multicolumn{2}{c}{I=2} \\ \hline
\multicolumn{1}{c|}{ENERGY} &
\multicolumn{1}{c}{LHS} &
\multicolumn{1}{c|}{RHS} &
\multicolumn{1}{c}{LHS} &
\multicolumn{1}{c|}{RHS} &
\multicolumn{1}{c}{LHS} &
\multicolumn{1}{c}{RHS} \\ \hline
      300. &    0.342 &     0.344 &     0.005 &     0.006 &    -0.029
&    -0.029 \\
      320. &    0.377 &     0.376 &     0.014 &     0.012 &    -0.043
&    -0.041 \\
      340. &    0.414 &     0.411 &     0.022 &     0.019 &    -0.055
&    -0.052 \\
      360. &    0.447 &     0.445 &     0.028 &     0.027 &    -0.067
&    -0.064 \\
      380. &    0.479 &     0.477 &     0.039 &     0.037 &    -0.080
&    -0.076 \\
      400. &    0.509 &     0.507 &     0.049 &     0.047 &    -0.090
&    -0.087 \\
      420. &    0.534 &     0.534 &     0.058 &     0.058 &    -0.100
&    -0.099 \\
      440. &    0.557 &     0.556 &     0.072 &     0.071 &    -0.113
&    -0.110 \\
      460. &    0.573 &     0.572 &     0.088 &     0.086 &    -0.122
&    -0.121 \\
      480. &    0.584 &     0.585 &     0.105 &     0.103 &    -0.132
&    -0.132 \\
      500. &    0.590 &     0.591 &     0.123 &     0.122 &    -0.142
&    -0.143 \\
      520. &    0.590 &     0.591 &     0.146 &     0.144 &    -0.152
&    -0.153 \\
      540. &    0.585 &     0.586 &     0.171 &     0.170 &    -0.161
&    -0.163 \\
      560. &    0.574 &     0.576 &     0.199 &     0.199 &    -0.171
&    -0.172 \\
      580. &    0.558 &     0.558 &     0.234 &     0.233 &    -0.178
&    -0.181 \\
      600. &    0.537 &     0.538 &     0.272 &     0.273 &    -0.186
&    -0.189 \\
      620. &    0.512 &     0.513 &     0.318 &     0.319 &    -0.194
&    -0.196 \\
      640. &    0.483 &     0.484 &     0.371 &     0.371 &    -0.201
&    -0.203 \\
      660. &    0.450 &     0.450 &     0.428 &     0.429 &    -0.209
&    -0.209 \\
      680. &    0.414 &     0.413 &     0.486 &     0.490 &    -0.214
&    -0.214 \\
      700. &    0.375 &     0.373 &     0.532 &     0.531 &    -0.222
&    -0.217 \\
\end{tabular}
\end  {table}

\narrowtext
\begin{table}
\caption{Data from energy-independent analysis
of Ochs\protect\tablenotemark[1]}\label{ochsdata}

\begin{tabular}{cc}
\multicolumn{1}{c}{Energy (MeV)} &
\multicolumn{1}{c}{$\delta_0$ (degrees)} \\
\tableline
610 &56.3 $\pm$ 3.2 \\
630 &59.5 $\pm$       2.9 \\
650 &65.6 $\pm$       3.2 \\
670 &62.5 $\pm$       3.5 \\
690 &68.8 $\pm$       3.6 \\
710 &74.5 $\pm$       3.8 \\
730 &79.4 $\pm$       3.6 \\
750 &81.2 $\pm$       5.7 \\
770 &79.9 $\pm$       3.9 \\
790 &77.5 $\pm$       5.7 \\
810 &84.1 $\pm$       3.3 \\
830 &84.4 $\pm$       2.6 \\
850 &87.1 $\pm$       2.5 \\
870 &89.2 $\pm$       2.5 \\
890 &93.2 $\pm$       2.9 \\
910 &103.3$\pm$       3.2 \\
\end{tabular}
\tablenotetext[1]{Ochs, Ref.\ \protect\cite{ochs}}
\end{table}

\mediumtext
\begin{table}
\caption{Comparison of the left and right hand sides of Eqs.(3.7a,b),
using phase shifts of Schenk, solution B}\label{schenkb}

\begin{tabular}{c|cc|cc|cc}\hline
\multicolumn{1}{c}{$~$} &
\multicolumn{2}{c|}{I=0} &
\multicolumn{2}{c|}{I=1} &
\multicolumn{2}{c}{I=2} \\ \hline
\multicolumn{1}{c|}{ENERGY} &
\multicolumn{1}{c}{LHS} &
\multicolumn{1}{c|}{RHS} &
\multicolumn{1}{c}{LHS} &
\multicolumn{1}{c|}{RHS} &
\multicolumn{1}{c}{LHS} &
\multicolumn{1}{c}{RHS} \\ \hline
      300. &    0.236 &     0.234 &     0.006 &     0.005 &    -0.053 &
-0.052 \\
      320. &    0.274 &     0.272 &     0.012 &     0.011 &    -0.064 &
-0.063 \\
      340. &    0.314 &     0.312 &     0.019 &     0.018 &    -0.075 &
-0.074 \\
      360. &    0.356 &     0.355 &     0.027 &     0.026 &    -0.087 &
-0.085 \\
      380. &    0.398 &     0.397 &     0.035 &     0.034 &    -0.098 &
-0.097 \\
      400. &    0.439 &     0.440 &     0.045 &     0.044 &    -0.109 &
-0.108 \\
      420. &    0.479 &     0.479 &     0.056 &     0.055 &    -0.121 &
-0.119 \\
      440. &    0.515 &     0.515 &     0.068 &     0.067 &    -0.132 &
-0.130 \\
      460. &    0.545 &     0.546 &     0.083 &     0.081 &    -0.143 &
-0.142 \\
      480. &    0.569 &     0.569 &     0.099 &     0.098 &    -0.153 &
-0.153 \\
      500. &    0.584 &     0.584 &     0.117 &     0.116 &    -0.164 &
-0.164 \\
      520. &    0.589 &     0.590 &     0.138 &     0.138 &    -0.174 &
-0.174 \\
      540. &    0.585 &     0.585 &     0.163 &     0.163 &    -0.184 &
-0.185 \\
      560. &    0.571 &     0.571 &     0.192 &     0.192 &    -0.194 &
-0.195 \\
      580. &    0.548 &     0.547 &     0.226 &     0.227 &    -0.203 &
-0.204 \\
      600. &    0.517 &     0.517 &     0.267 &     0.268 &    -0.212 &
-0.213 \\
      620. &    0.480 &     0.479 &     0.315 &     0.316 &    -0.221 &
-0.222 \\
      640. &    0.439 &     0.439 &     0.370 &     0.372 &    -0.229 &
-0.230 \\
      660. &    0.395 &     0.394 &     0.433 &     0.437 &    -0.237 &
-0.237 \\
      680. &    0.350 &     0.350 &     0.496 &     0.495 &    -0.244 &
-0.244 \\
      700. &    0.304 &     0.305 &     0.540 &     0.538 &    -0.252 &
-0.249 \\
\end{tabular}
\end  {table}

\begin{table}
\caption{Analysis of $\phi_0$ based on the phase shift data of Ochs
and of Estabrooks and Martin (E-M), extrapolated to threshold, for
fixed values of the scattering length $a_0$, using the parametrization
of Schenk.}\label{phi0fits}
\begin{tabular}{ccccccccc}
\multicolumn{1}{c||}{Data} &
\multicolumn{1}{c}{$b_0$}&
\multicolumn{1}{c}{$c_0$}&
\multicolumn{1}{c}{$E_0$}&
\multicolumn{1}{c||}{$\chi^2$/d.f.}&
\multicolumn{1}{c}{$\alpha/\beta$}&
\multicolumn{1}{c}{$\tau_0$}&
\multicolumn{1}{c||}{$\chi^2$/d.f.}&
\multicolumn{1}{c}{$\chi^2$/d.f.}\\
\multicolumn{1}{c||}{$~$}&
\multicolumn{4}{c||}{$~$}&
\multicolumn{2}{c}{$~$}&
\multicolumn{1}{c||}{(imprvd.)}&
\multicolumn{1}{c}  {(stand.)}\\
\hline
\multicolumn{9}{c}{$a_0$ = 0.20} \\
\hline
\multicolumn{1}{c||}{Ochs ``a''} &
\multicolumn{1}{c}{0.393}&
\multicolumn{1}{c}{-0.0356} &
\multicolumn{1}{c}{867.8} &
\multicolumn{1}{c||}{11/13} &
\multicolumn{3}{c||}{~} &
\multicolumn{1}{c}{~} \\
\multicolumn{1}{c||}{Ochs ``b''} &
\multicolumn{1}{c}{0.348}&
\multicolumn{1}{c}{-0.0292} &
\multicolumn{1}{c}{863.3} &
\multicolumn{1}{c||}{10/13} &
\multicolumn{1}{c}{2.32(4)} &
\multicolumn{1}{c}{-0.094(6)} &
\multicolumn{1}{c||}{67/63} &
\multicolumn{1}{c}{690/63} \\
\multicolumn{1}{c||}{Ochs ``c''} &
\multicolumn{1}{c}{0.298}&
\multicolumn{1}{c}{-0.0206} &
\multicolumn{1}{c}{858.8} &
\multicolumn{1}{c||}{11/13} &
\multicolumn{3}{c||}{~} &
\multicolumn{1}{c}{~} \\ \hline
\multicolumn{1}{c||}{E-M ``a''} &
\multicolumn{1}{c}{0.253} &
\multicolumn{1}{c}{-0.0184} &
\multicolumn{1}{c}{818.6} &
\multicolumn{1}{c||}{15/15} &
\multicolumn{3}{c||}{~} &
\multicolumn{1}{c}{~} \\
\multicolumn{1}{c||}{E-M ``b''} &
\multicolumn{1}{c}{0.229} &
\multicolumn{1}{c}{-0.0147} &
\multicolumn{1}{c}{814.0} &
\multicolumn{1}{c||}{14/15} &
\multicolumn{1}{c}{1.581(3)} &
\multicolumn{1}{c}{-0.236(5)} &
\multicolumn{1}{c||}{30/63} &
\multicolumn{1}{c}{677/63}\\
\multicolumn{1}{c||}{E-M ``c''} &
\multicolumn{1}{c}{0.205} &
\multicolumn{1}{c}{-0.0109} &
\multicolumn{1}{c}{809.6} &
\multicolumn{1}{c||}{15/15} &
\multicolumn{3}{c||}{~} &
\multicolumn{1}{c}{~} \\ \hline
\multicolumn{9}{c}{$a_0$ = 0.26} \\ \hline
\multicolumn{1}{c||}{Ochs ``a''} &
\multicolumn{1}{c}{0.369} &
\multicolumn{1}{c}{-0.0339} &
\multicolumn{1}{c}{867.4} &
\multicolumn{1}{c||}{11/13} &
\multicolumn{3}{c||}{~} &
\multicolumn{1}{c}{~} \\
\multicolumn{1}{c||}{Ochs ``b''} &
\multicolumn{1}{c}{0.324} &
\multicolumn{1}{c}{-0.0274} &
\multicolumn{1}{c}{863.1} &
\multicolumn{1}{c||}{10/13} &
\multicolumn{1}{c}{3.50(3)} &
\multicolumn{1}{c}{-0.154(6)} &
\multicolumn{1}{c||}{48/63} &
\multicolumn{1}{c}{2328/63} \\
\multicolumn{1}{c||}{Ochs ``c''} &
\multicolumn{1}{c}{0.274} &
\multicolumn{1}{c}{-0.0188} &
\multicolumn{1}{c}{858.6} &
\multicolumn{1}{c||}{11/13} &
\multicolumn{3}{c||}{~} &
\multicolumn{1}{c}{~} \\ \hline
\multicolumn{1}{c||}{E-M ``a''} &
\multicolumn{1}{c}{0.227} &
\multicolumn{1}{c}{-0.0163} &
\multicolumn{1}{c}{817.8} &
\multicolumn{1}{c||}{14/15} &
\multicolumn{3}{c||}{~} &
\multicolumn{1}{c}{~} \\
\multicolumn{1}{c||}{E-M ``b''} &
\multicolumn{1}{c}{0.203} &
\multicolumn{1}{c}{-0.0126} &
\multicolumn{1}{c}{813.3} &
\multicolumn{1}{c||}{13/15} &
\multicolumn{1}{c}{2.86(3)} &
\multicolumn{1}{c}{-0.305(2)} &
\multicolumn{1}{c||}{73/63} &
\multicolumn{1}{c}{4947/63} \\
\multicolumn{1}{c||}{E-M ``c''} &
\multicolumn{1}{c}{0.179} &
\multicolumn{1}{c}{-0.0087} &
\multicolumn{1}{c}{808.9} &
\multicolumn{1}{c||}{14/15} &
\multicolumn{3}{c||}{~} &
\multicolumn{1}{c}{~} \\
\end{tabular}
\end{table}

\begin{references}
\bibitem{sw79}S. Weinberg, Physica {\bf A96}, 327 (1979).
\bibitem{gl84}J. Gasser and H. Leutwyler, Ann.\ Phys.\ (NY)\ {\bf
158}, 142 (1984).
\bibitem{gl85}J. Gasser and H. Leutwyler, Nucl.\ Phys.\ {\bf B250},
465 (1985).
\bibitem{drv}J. Donoghue, C. Ramirez and G. Valencia, Phys.\ Rev. {\bf
D38}, 2195 (1988).
\bibitem{gor-gw}M. Gell-Mann, R.J. Oakes and B. Renner, Phys.\ Rev.
{\bf 175}, 2195 (1968); S. Glashow and S. Weinberg, Phys.\  Rev.\
Lett.\ {\bf 20}, 224 (1968).
\bibitem{fss91}N. H. Fuchs, H. Sazdjian and J. Stern, Phys.\ Lett.\ B
{\bf269}, 183 (1991).
\bibitem{sumrules}S. Weinberg, Phys.\ Rev.\ Lett.\ {\bf 18}, 188
(1967); T. Das, V. Mathur and S. Okubo, Phys.\ Rev.\ Lett.\ {\bf 18},
781 (1967); G. Ecker, J. Gasser, H. Leutwyler, A. Pich and E. De
Rafael, Phys.\ Lett.\ B {\bf 223}, 425 (1989); G. Ecker, J. Gasser, A.
Pich and E. De Rafael, Nucl.\ Phys.\ {\bf B321}, 311 (1989); H.
Leutwyler, Nucl.\ Phys.\ {\bf B337}, 108 (1990); J. F. Donoghue and B.
R. Holstein, Phys.\ Rev.\ {\bf D46}, 4076 (1992).
\bibitem{njl}Y. Nambu and G. Jona-Lasinio, Phys. Rev. {\bf 122}, 345
(1961); {\bf 124}, 246 (1961).
\bibitem{w-gl}S. Weinberg, in {\it A Festschrift for I.I. Rabi}, ed.
L. Motz (New York Academy of Sciences, New York, 1977) p. 185; J.
Gasser and H. Leutwyler, Phys.\ Rep.\ {\bf 87}, 77 (1982).
\bibitem{fss90}N. H. Fuchs, H. Sazdjian and J. Stern, Phys.\ Lett.\ B
{\bf238}, 380 (1990).
\bibitem{koch-piet}R. Koch and E. Pietarinen, Nucl.\ Phys.\ {\bf
A336}, 331 (1980); see also R. Koch and M. Hutt, Z.\ Phys.\ C {\bf 19},
119 (1983).
\bibitem{arndt-etal}R.A. Arndt {\it et al.}, Phys.\ Rev.\ Lett.\ {\bf
65}, 157 (1990).
\bibitem{gammagamma}J. Bijnens and F. Cornet, Nucl.\ Phys.\ {\bf B296},
557 (1988); J. F. Donoghue, B. R. Holstein and Y. C. Lin, Phys.\ Rev.\
{\bf D37}, 2423 (1988); H. Marsiske {\it et al.}, Phys.\ Rev.\ {\bf
D41}, 3324 (1990); D. Morgan and M. R. Pennington, Phys.\ Lett.\ B
{\bf 272}, 134 (1992); M. R. Pennington, in {\it Proceedings of the
Workshop on Physics and Detectors for DAPHNE, The Frascati Phi
Factory}, ed. G. Pancheri (Frascati, Lab. Naz.  Frascati, 1991), p.
379 (1992).
\bibitem{eta3pi}J. Gasser and H. Leutwyler, Nucl.\ Phys.\ {\bf B250},
539 (1985).
\bibitem{sw66}S. Weinberg, Phys.\ Rev.\ Lett.\ {\bf 17}, 616 (1966).
\bibitem{yellow}J. L. Petersen, {\it The $\pi\pi$ interaction},
lectures given in the Academic Training Programme (CERN, 1975-6), CERN
report 77-04 (unpublished).
\bibitem{fp}C. D. Froggatt and J. L. Petersen, Nucl.\ Phys.\ {\bf
B129}, 89 (1977).
\bibitem{nagels}M. M. Nagels {\it et al.}, Nucl.\ Phys.\ {\bf B147},
189 (1979); O. Dumbrajs {\it et al.}, Nucl.\ Phys.\ {\bf B216}, 277
(1983).
\bibitem{donoghuereview}J. Donoghue, Ann. Rev. Nucl. Part. Phys. {\bf
39}, 1 (1989).
\bibitem{roy}S. M. Roy, Phys.\ Lett.\ {\bf 36B}, 353 (1971); Helv.\
Phys.\ Acta {\bf 63}, 627 (1990); J. L. Basdevant, J. C. Le Guillou
and H. Navelet, Nuovo Cim.\ {\bf 7A}, 363 (1972).
\bibitem{edd}A. S. Eddington, Proc. Roy. Soc. {\bf A122}, 353 (1929).
\bibitem{chew-mandelstam}G. F. Chew and S. Mandelstam, Phys.\ Rev.\
{\bf 119}, 467 (1960).
\bibitem{bfp}J. L. Basdevant, C. D. Froggatt and J. L. Petersen,
Phys.\ Lett.\ B {\bf 41}, 173, 178 (1972); Nucl.\ Phys.\ {\bf B72},
413 (1974); M. R. Pennington and S. D. Protopopescu, Phys.\ Rev.\ {\bf
D7}, 1429, 2591 (1973).
\bibitem{iterate}D. Atkinson and T. P. Pool, Nucl.\ Phys.\ {\bf 81B},
502 (1974); T. P. Pool, Nuovo Cim.\ {\bf 45A}, 207 (1978); A. C.
Heemskerk and T. P. Pool, Nuovo Cim. {\bf 49A}, 393 (1979).
\bibitem{ochs-newsletter}W. Ochs, $\pi-N$ Newsletter {\bf 3}, 25
(1991).
\bibitem{gm}J. Gasser and U.-G. Meissner, Phys.\ Lett.\ B {\bf 258},
219 (1991).
\bibitem{em74}P. Estabrooks and A. D. Martin, Nucl.\ Phys.\ {\bf B79},
301 (1974).
\bibitem{cern-munich}B. Hyams {\it et al.}, Nucl.\ Phys.\ {\bf B64},
134 (1973); W. Hoogland {\it et al.}, Nucl.\ Phys.\ {\bf B69}, 266
(1974). .
\bibitem{ke4}L. Rosselet {\it et al.}, Phys.\ Rev.\ D {\bf 15}, 574
(1977).
\bibitem{schenk}A. Schenk, Nucl.\ Phys.\ {\bf B363}, 97 (1991).
\bibitem{ochs}W. Ochs, Thesis, Ludwig-Maximilians-Universit\"at,
1973.
\bibitem{mass-dimension}N. H. Fuchs, Nuovo Cim.\ Lett.\ {\bf 27}, 21
(1980); J. Stern, Black Forest meeting on quark masses and lattice
gauge theories (Todtnauberg, 1982), unpublished; J.R. Crewther, Phys.
Lett. B{\bf 176}, 172 (1984); G.A. Christos, Austr. J. Phys. {\bf 39},
347 (1986).
\bibitem{rigg}C. Riggenbach, J. Gasser, J. F. Donoghue and B. R.
Holstein, Phys.\ Rev.\ {\bf D43}, 127 (1991).
\end{references}
\end{document}